\documentclass[epj]{svjour}

\usepackage{graphics}
\usepackage{epsfig}

\journalname{The European Physical Journal E}

\newcommand{\dy}{\raisebox{1.6ex}{\rotatebox{180}{\textsf{Y}}}}
\newcommand{\uy}{\textsf{Y}}
\newcommand{\nh}{\hat{\vec{n}}}
\newcommand{\uh}{\hat{\vec{u}}}
\newcommand{\vh}{\hat{\vec{v}}}
\newcommand{\ev}{\vec{e}}
\newcommand{\fv}{\vec{f}}
\newcommand{\pv}{\vec{p}}
\newcommand{\qv}{\vec{q}}
\newcommand{\rv}{\vec{r}}
\newcommand{\Cv}{\vec{C}}
\newcommand{\Ev}{\vec{E}}
\newcommand{\Fv}{\vec{F}}
\newcommand{\Jv}{\vec{J}}

\newcommand{\Qv}{\vec{Q}}
\newcommand{\Sv}{\vec{S}}
\newcommand{\Xv}{\vec{X}}
\newcommand{\Mm}{\tens{M}}
\newcommand{\eqname}{Boltzmann }
\newcommand{\Eqname}{Boltzmann }
\newlength{\smallfig}

\begin{document}

\title{
Directed force chain networks and stress response in static granular materials
}
\author{
J.~E.~S.~Socolar\inst{1} \and D.~G.~Schaeffer\inst{2} \and P.~Claudin\inst{3}
}

\institute
{Department of Physics and Center for Nonlinear and Complex Systems, Duke University, Durham, NC 27708, USA
\and
Department of Mathematics and Center for Nonlinear and Complex Systems, Duke University, Durham, NC 27708, USA
\and
Laboratoire des Milieux D\'{e}sordonn\'{e}s et H\'{e}t\'{e}rog\`{e}nes (UMR 7603), 4 place Jussieu - case 86, 75252 Paris Cedex 05, France
}

\date{\today}

\abstract{
  A theory of stress fields in two-dimensional granular materials
  based on directed force chain networks is presented.  A general
  \eqname equation for the densities of force chains in different
  directions is proposed and a complete solution is obtained for a
  special case in which chains lie along a discrete set of directions.
  The analysis and results demonstrate the necessity of including
  nonlinear terms in the \eqname equation.  A line of nontrivial fixed
  point solutions is shown to govern the properties of large systems.
  In the vicinity of a generic fixed point, the response to a
  localized load shows a crossover from a single, centered peak at
  intermediate depths to two propagating peaks at large depths that
  broaden diffusively.
}

\authorrunning{Socolar, Schaeffer, and Claudin}
\titlerunning{Directed force chain networks and stress response in static granular materials}
\maketitle


\section{Introduction}
\label{sec:intro}

The response to a localized force applied at the boundary of a
semi-infinite sample is an essential feature of any macroscopic
material.  For materials that are well described by linear elasticity
theory, the response can be calculated by standard techniques: using
the relation of the stresses to a displacement field, one constructs
partial differential equations for components of the stress tensor
$\sigma$, then solves them with appropriate boundary conditions on
$\sigma$ and its derivatives.  For a wide class of granular materials,
however, the absence of an energy expressible in terms of microscopic
(or grain scale) displacements leads to serious difficulties in
deriving the stress response.  

Numerous attempts have been made, many quite recent, to close the
system of stress equilibrium equations by deriving (or simply
guessing) a relation between the components of the stress tensor
beyond those required by Newton's laws.  The needed relation may be
extracted from assumptions concerning yield thresholds in
elastoplastic theories \cite{nedderman,cantelaubePG}, from an analysis
of physics at the grain scale,\cite{edwards}, from general symmetry
principles and considerations of simplicity \cite{bcc,cbcw}, or from
considerations applicable to isostatic networks \cite{witten,ball}.  A
special case that has received much attention is that of frictionless,
circular disks.\cite{roux,head} Alternatively, lattice models for the
configuration of individual inter-grain forces can serve as the basis
for the derivation of average stress patterns while simultaneously
addressing the statistical properties of forces at the grain scale.
\cite{bcc,qmodel1,socolar,eloy}

In this paper we develop a theory of stress distribution in
noncohesive granular materials based on the physics of {\em force
  chains}, rather than macroscopic stresses or grain scale
interactions.  Experimental probes and numerical computations of
stress patterns in granular materials universally show filamentary
networks that support strong forces between grains.
\cite{howellPG,radjai,cwbc} We take this as motivation to describe all
of the forces in the system in terms of linear ``force chains'' that
combine to form a network whose large scale properties determine the
macroscopic stress field.  To describe such networks, Bouchaud et al.\ 
recently introduced the ``\dy\uy -model'' (pronounced ``double Y
model'').\cite{bcol} Here we develop the \dy\uy-model in greater
detail, point out certain essential ways in which the original
formulation and analysis was incomplete, and calculate stress patterns
and response functions in special cases.

In the \dy\uy-model, stresses are modeled as networks of interacting
line segments that carry compressive forces.  A crucial feature of the
model is that the force chains are {\em directed}: each is conceived
as ``propagating'' from one of its endpoints to the other, in a sense
that will be clarified below.  Chains are created by three processes:
(1) imposed conditions at a boundary; (2) the splitting of one chain
into $D$ in a manner that preserves force balance at the splitting
point, where $D$ is the spatial dimension of the material; or (3) the
fusion of $D$ chains into one.  They then propagate until they are
destroyed via splitting (``\dy'' processes), fusion (``\uy''
processes), or traversal of a sample boundary.  We refer to a static
network of chains interacting in this way as a ``directed force chain
network'' (DFCN).

Our goal is to determine the large scale structure of DFCN's and their
average responses to small, local perturbations at a boundary.  We
first develop a \eqname equation governing the spatial variations in
the densities of chains supporting certain force intensities and
oriented in particular directions.\footnote{We use the term
  ``Boltzmann equation'' loosely.  It does not imply the existence of
  a causal structure or H-theorem.} We then seek solutions in the
special case where the chains are restricted to lie along a discrete
set of directions.  The results highlight several subtle features of
the directed force chain system, and show that the response of such
systems can have a rich structure that includes some surprising
effects.

Our primary result can best be stated with reference to the results
previously obtained via standard elasticity theory or posited closure
relations.  In standard elasticity theories of two-dimensional
isotropic systems, the response to a normal force applied at one point
($x=0$) of a half plane is \cite{LLelasticity}
\begin{equation}
\sigma_{zz} = \frac{2}{\pi}\frac{z^3}{\left(x^2+z^2\right)^2},
\end{equation}
where $z=0$ at the surface and increases with depth.  Thus at any
fixed depth $z$, the response $\sigma_{zz}$ consists of a single peak
centered on $x=0$, with a width that grows linearly with depth.
Though $\sigma_{zz}$ for anisotropic materials can have two peaks, the
response in this geometry is always invariant under uniform
dilations and rescaling: $(x\rightarrow a x, z\rightarrow a z)$
yields $\sigma_{ij}\rightarrow (1/a)\sigma_{ij}$.

In another class of theories based on closure relations that make no
reference to a displacement field, the stress equilibrium equations
are hyperbolic.  In such cases, the response function consists of two
peaks that propagate linearly away from $x=0$.  The inclusion of weak
disorder in these models leads to a diffusive broadening of the peaks
with increasing depth.\cite{cbcw} Note that this form of response
differs markedly from the elastic case, as the ratio of peak width to
propagation distance decreases like $1/\sqrt{z}$ at large depths
rather than remaining constant.  Perhaps even more importantly, the
two types of theory require different manners of specifying boundary
conditions.  Whereas standard elasticity theory requires the
specification of two conditions on the stresses (and/or their
derivatives) at all points on the boundary, the hyperbolic models
require (permit) the specification of stresses on only the top
boundary.

In the directed force chains system, we find that both top and bottom
boundary conditions are important, but the solutions and response
functions in deep systems have a richer structure than either the
elastic or hyperbolic theories suggest.  For very shallow depths
beneath the surface, the response may have one or two peaks, depending
on the details of the applied forces.  For intermediate depths, up to
several times the average length of a force chain, the response has a
single peak and may appear quite similar to a standard elastic
response.  For large depths, however, two peaks emerge that grow
diffusively with depth.  This latter behavior suggests that at the
largest scales a hyperbolic model for stress equilibrium may be
appropriate even for systems with strong disorder, though important
questions remain concerning the role of the discrete set of directions
in sustaining the two-peaked structure.

A second result to be emphasized is the failure of the linearized
theory obtained by assuming that all chain densities are small.
Formally, this assumption is equivalent to neglecting the fusion of
chains, keeping only those terms in the \eqname equation that describe
splitting events.  For intermediate and deep systems, internal
consistency of the theory requires that the response be computed by
linearizing around a nontrivial fixed point of the \eqname equation.
This sheds new light on previous efforts to derive a linear elasticity
theory of DFCN's.  It also has important implications for the
numerical modeling of DFCN's and the interpretation of experimental
measurements of stress response in granular systems.

In addition to these two central results, we point out several curious
features of DFCN's in a simple slab geometry in two dimensions.  These
include an effective Poisson ratio that may depend on the way in which
a load is applied, an exponentially localized region of increased
horizontal stresses in the middle of a slab subject to horizontally
uniform vertical stresses at top and bottom, and a distinction between
responses to identical applied macroscopic stresses with different
distributions of force chain densities in various directions.

The paper is organized as follows.  In the Section~\ref{sec:master},
we introduce the concept of the directed force chain, define the
constants, variables and functions that enter our theory, develop a
\eqname equation that governs spatial variations of the densities of
force chains, and discuss the boundary conditions on the equation.  We
then define a special case of the model in which chain directions are
restricted to discrete set that is amenable to analytical treatment.
In Section~\ref{sec:uniform}, we solve the discrete \eqname equation
in a slab of depth $z$ for the case of applied loads that are uniform
in the horizontal direction, identifying the (nontrivial) solutions
relevant for the computation of response functions.  In
Section~\ref{sec:response}, we analyze the response (in the discrete
model) in the vicinity of a generic fixed point, present some results
for other choices of discrete directions in 2D and describe a possible
generalization to 3D.  We also discuss the relation of our analysis
methods to the long-wavelength theory presented in Ref.~\cite{bcol}.
We conclude, in Section~\ref{sec:conclusion} with discussions of the
issue of numerical generation of DFCN's and some open questions.

\section{\Eqname equation for directed force chains}
\label{sec:master}
\subsection{Definitions, the \eqname equation, and its boundary conditions}
\label{sec:chaindef}
\begin{figure}
\begin{center}
\epsfxsize=\linewidth
\epsfbox{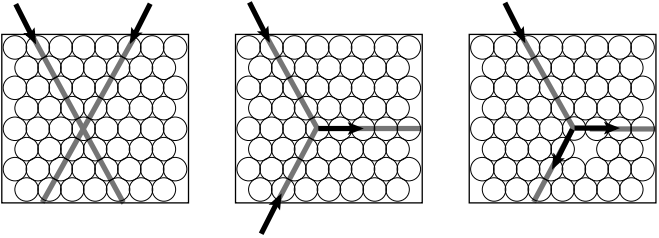}
\caption{Illustration of force chains.  (a) Two force chains
  initiated at the top surface cross without interacting.  (b) When
  one of the force chains is initiated at the bottom, a fusion becomes
  possible.  (c) A force chain initiated at the top that splits upon
  encountering a defect.
\label{fig:forcechains}}
\end{center}
\end{figure}

A {\em force chain} is defined as a line segment that supports a
compressive force.  The {\em direction} of a force chain has two
features.  First, the segment makes a particular angle with the
vertical.  Second, the two ends of the segment are distinguished --
the segment has a ``beginning'' and an ``end'' determined by the role
the chain plays in the network.  Fig.~\ref{fig:forcechains}
illustrates the latter point.  The figure shows a two-dimensional
system consisting of a hexagonal array of grains.  In panel (a), two
forces are applied at the top boundary and the forces propagate along
chains, crossing at the central grain.  In panel (b), a similar
situation is shown, but this time one of the chains is assumed to be
specified by fixing its position at the bottom boundary instead of the
top.  In this case, a fusion of the two chains at the central grain is
permitted (though not required) and the resulting configuration may be
different, demonstrating the inequivalence of the two possible choices
of direction for the chain whose boundary condition was changed.  The
horizontal chain is assumed to ``begin'' at the fusion point and
``end'' at the boundary.

Thus each chain has an intensity, $f$, and a direction $\nh$.  Though
we will sometimes combine the two and express them as a vector $\fv$,
it is important to keep in mind that the compressive force supported
by the chain has no forward or backward direction.  A grain that is
part of a force chain experiences no net force, only a stress with its
major principal axis along the chain.  The distinction between the
forward and backward directions of the chain refers to its own
boundary conditions, which are ultimately related to the boundary
conditions at the edge of the sample, though this relation may be
difficult to determine.

In addition to fusion of chains, it is possible for a chain to split
into two (in two dimensions).  Fig.~\ref{fig:forcechains}(c) shows a
splitting event induced by the presence of a defect in the system.  In
general, ordered regions and defects will not be so easy to identify
by eye, and all sites are assumed to produce splittings and permit
fusions with probabilities drawn from a specified angular
distribution, as described below.

\begin{figure}
\begin{center}
\epsfxsize=\linewidth
\epsfbox{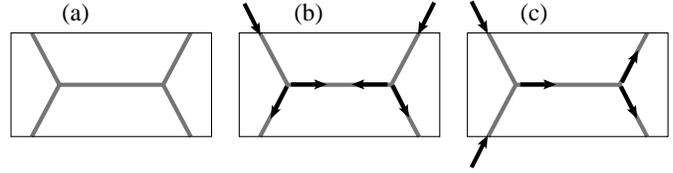}
\caption{(a) A seemingly ambiguous configuration of force chains.  
  (b) Boundary conditions that imply an annihilation event.
  (c) Boundary conditions that give a unique direction to the middle chain.
\label{fig:annihilation}}
\end{center}
\end{figure}

To clarify the way in which directions may be assigned to force chains, it
may be helpful to consider the configuration shown in
Fig.~\ref{fig:annihilation}.  Here the question arises as to how one
can assign a direction to the chain in the middle of panel (a).  If
the situation is as shown in (b), with boundary imposed at the two
sites on the top surface, then the middle chain must be interpreted as
consisting of two oppositely directed chains that annihilate in the
middle.  The same is true for any other configuration of boundary
conditions that force this configuration on the middle chain.
Alternatively, there exist other possible specifications of the
boundary conditions that give the middle chain a specific orientation,
as shown in (c).  In developing the theories below, we neglect all
annihilations of the type shown in (b).  The reason is that for very
narrow force chains, the probability of two chains meeting head on is
very low.  In granular systems, it may be argued that the force chain
widths are effectively of the order of a grain size, so that
annihilation events are not entirely negligible.  For the present
paper, we neglect effects associated with the finite sizes of grains.
Preliminary investigations strongly suggest that inclusion of terms
corresponding to annihilations does not change the general features of
the results.

For clarity of presentation, from here on we work in two dimensions
unless explicitly noted otherwise.  The main difference in $D$
dimensions is that vertices where chains fuse or split generically
require $D+1$ chains in order to achieve force balance, since force
balance among fewer chains would require the highly improbable event
that one of them lie in the subspace spanned by the others.

We begin with the assumption that a 2D granular material can be
thought of as a collection of local environments through which single
force chains are either transmitted or split into two, and pairs of
intersecting force chains either pass through each other or fuse into
one.  We also assume that the force chains carry all of the stresses
in the system.  Newton's laws are built into the model via the
constraint that the forces at each splitting and fusion event must
balance.  In this formulation of the model, there are never any
unbalanced torques, as the three chains associated with a splitting or
fusion always meet at a single point.

Following the suggestion of Ref.~\cite{bcol}, we construct a \eqname
equation governing the densities of chains as follows.  Let
$P(f,\nh,\rv)$ represent the probability of finding a force chain of
intensity $f$ in direction of unit vector $\nh$ passing through the
spatial point $\rv$.  In other words, $\int_{g}^{g+\delta}\!df \int\!
d\nh\;|\nh\!\cdot\!\uh|P(f,\nh,\rv)$ is the number of chains of
intensity between $g$ and $g+\delta$ that cross a unit length line
segment through $\rv$ and perpendicular to $\uh$.  Note that $P$ has
dimensions of $1/(force\times length)$ in 2D and cannot be negative.
The local stress tensor, expressed in terms of the coarse grained
quantity $P$, is given by \cite{bcol}
\begin{equation}\label{eqn:stresstensor}
\sigma_{\alpha\beta} = \int_{0}^{\infty}\!\!df \int\!\!d\Omega\;
n_{\alpha}n_{\beta} f P(f,\nh,\rv).
\end{equation}
Here $\Omega$ represents the $D$-dimensional angular direction of
$\nh$ divided by the full solid angle so that $\int\!d\Omega=1$.
For our 2D discussion, we will use $\theta\in(-\pi,\pi]$ to designate the
direction $\nh$, and $d\Omega = d\theta/2\pi$.

We define two functions $\phi_s$ and $\phi_f$ that characterize the
probability of generating chains at various angles when a splitting or
fusion occurs.  In the most general case, these may be functions of
the intensities and directions of all three chains, but must include
delta functions ensuring force balance and Heaviside functions
$\Theta(f)$ ensuring that all force intensities are positive (all
force chains in a noncohesive material must carry compressive stress,
not tensile).
\begin{eqnarray}
\phi_{a}(\fv_1\,|\,\fv_2,\fv_3) & = & 
\delta(\fv_2+\fv_3-\fv_1)\;\Theta(f_1)\,\Theta(f_2)\,\Theta(f_3) \nonumber\\
\ & \ & \times\; \psi_{a}(\theta_1\,|\,\theta_2,\theta_3)\,|\sin(\theta_2-\theta_3)|, 
\label{eqn:phi}
\end{eqnarray}
where $\fv$ indicates the pair $(f,\nh)$ and $\psi_{a}$ is normalized
such that $\int\! d\theta_2 d\theta_3 df_2 df_3\,\phi_{a}=1$.  The vertical
bar is used in expressing the arguments of $\phi_{a}$ and $\psi_{a}$
to indicate that the first argument is associated with the minus sign
in the delta function.  For splittings, the first argument corresponds
to the single incoming chain.  For fusions, it corresponds to the
single outgoing chain.  

The explicit factor of $|\sin(\theta_2-\theta_3)|$ in Eq.~(\ref{eqn:phi})
serves several purposes.  For the splitting function, it is included
to remind us that $\phi_s$ must vanish for splitting events in which
the outgoing chains are collinear and oppositely directed, as this
would generate infinite force intensities.  It is also reasonable to
assume that splitting events with outgoing chains close to the same
direction are rare, though $\psi_s$ could of course be chosen to make
such events as probable as desired.  For the fusion function, the
factor gives the density of intersection points per unit area for unit
chain densities of chains with orientation $\theta_2$ and $\theta_3$, which
should clearly affect the probability of fusions occuring.  This
factor also simplifies the normalization of $\psi_{a}$, since for
$\theta_1=0$ and any function $A$ we have
\begin{eqnarray}
\int_0^{\infty}&\!\!df_2&df_3\;\delta(\fv_2+\fv_3-\fv_1) A(f_1,f_2,f_3) \label{eqn:deltafunction} \\
\ & = & \int_0^{\infty}\!\!df_2\,df_3\, \delta(f_2\cos\theta_2+f_3\cos\theta_3-f_1) \nonumber \\
\ & \ & \quad\quad\times\; \delta(f_2\sin\theta_2+f_3\sin\theta_3) A(f_1,f_2,f_3) \nonumber \\
\ & = & \frac{1}{\sin(\theta_2-\theta_3)}
   A\left(f_1,\frac{f_1\sin\theta_2}{\sin(\theta_2-\theta_3)},\frac{f_1\sin\theta_3}{\sin(\theta_2-\theta_3)}\right).
 \nonumber
\end{eqnarray}

Note that $\psi_{a}$ must be symmetric in its second and third
arguments; in the case of splitting (fusion), switching the labels of
the two outgoing (incoming) chains cannot change the physics.  Note
also that the delta function is really the product of two delta
functions, as shown in Eq.~(\ref{eqn:deltafunction}), one for each
component of force balance, and therefore has a dimension of inverse
force squared.  We will restrict our attention to homogeneous and
isotropic systems, for which $\psi_{a}$ does not vary with position
and depends only upon the relative angles between its three arguments.

In general, $\psi_s$ and $\psi_f$ are different, owing to the physical
difference between splitting and fusion events.  As illustrated in
Fig.~\ref{fig:forcechains} for a particularly simple system, the
type of defect required to induce splitting is not necessarily present
at a fusion.  The boundary conditions on the various chains involved
make for different probabilities between splittings and fusions for a
given local geometry of the granular packing. 

The following \eqname equation describes the variation of
$P(f,\theta,\rv)$ along the direction $\nh$, where $\lambda$ and $\mu$
are constants required on dimensional grounds and, for notational
convenience, we have dropped the $\rv$ from the argument of all of the
$P$'s:
\begin{eqnarray}\label{eqn:master}
(\nh\cdot\nabla) & P & \!\!(\fv) = \nonumber \\
\ & \frac{1}{\lambda} & 
\left[ -P(\fv) + 2 \int\! df^{\prime}df^{\prime\prime}d\theta^{\prime}d\theta^{\prime\prime}
\;\phi_{s}(\fv^{\prime}\,|\,\fv,\fv^{\prime\prime})P(\fv^{\prime})\right] \nonumber \\
 + & \frac{\mu}{\lambda} &
\int\! df^{\prime}df^{\prime\prime}d\theta^{\prime}d\theta^{\prime\prime}
\big[ \phi_{f}(\fv\,|\,\fv^{\prime},\fv^{\prime\prime})P(\fv^{\prime})P(\fv^{\prime\prime}) \nonumber \\
\ & \  & \quad\quad\quad\quad\quad\quad - 2 \phi_{f}(\fv^{\prime}\,|\,\fv,\fv^{\prime\prime})P(\fv)P(\fv^{\prime\prime}) \big] .
\end{eqnarray} 
In this expression, all integrals over angles run from $-\pi$ to $\pi$
and all integrals over forces run from $0$ to $\infty$.  The first
term on the right hand side in Eq.~(\ref{eqn:master}) describes the
decay of force chain density in the direction of the chain due to
splitting of the chain.  $\lambda$ thus corresponds to the average
distance a chain propagates before splitting, in the absence of all
other chains.  In principle, $\lambda$ may be a function of the
direction $\nh$ and position, but is a single constant for homogeneous
and isotropic systems.  The second term on the right hand side
describes the increase in $P(\fv)$ due to the splitting of other
chains.  The factor of two in this term is due to the fact that
$\phi_{s}$ has been normalized to unity while a splitting event
produces two outgoing chains.  The final integral in
Eq.~(\ref{eqn:master}) is quadratic in $P$ and describes the effects
of fusions.  The quantity $\mu$ has dimensions of $1/P$, or
$force\times length$ in 2D.  As with $\lambda$, we assume $\mu$ is
independent of $\theta$.  The factor of $2$ in arises because the chain
in the $\theta$ direction can be either of the two incoming chains in a
fusion.

The boundary conditions to be applied to Eq.~(\ref{eqn:master}) are
suggested by Fig.~\ref{fig:forcechains}.  At each point on the
boundary of a sample we may specify the density of ingoing chains of
any intensity, but we may not specify the density of outgoing chains.
This means that there is no simple way to assign boundary conditions
corresponding to the overall stress on a boundary; the density of
outgoing chains, which add to the normal and shear stresses just as
ingoing ones do, is determined by the splittings and fusions within
the sample.  
In the slab geometry, however, it is possible to specify the total
vertical force applied to the top and bottom boundaries, due to the
symmetry of the force network under multiplication of all intensities
by a common factor. One
may first specify the density of inward directed chains, then compute
the density of outward directed chains using the \eqname equation,
compute the stresses at the boundary, then rescale all of the forces
in the system to arrive at the desired boundary stress.  Satisfying
conditions on both the normal and shear stress may also require
adjustment of the relative densities of chains in different
inward-pointing directions.  The manner in which boundary conditions
on the chain densities can be specified will be clarified by the
precise treatment of a special case in Section~\ref{sec:uniform}.

The fact that we may only specify the densities of ingoing chains at
a boundary is a part of the definition of the \dy\uy-model that deserves
further comment.  Microscopically, when we specify that a chain of a
given strength must pass through a certain point on the boundary, the
boundary condition {\em defines} the direction of that chain to be
inward.  This is physically plausible because we would not expect to
be able to apply a boundary condition that requires an intricate
conspiracy of fusions and splittings to create a chain that propagates
outward at the point in question.  In the context of the Boltzmann
equation, in which we do not specify individual chains but only
densities, this is reflected mathematically in the fact that
specification of outgoing chain densities leads to nonsense in certain
generic situations, as shown in Section ~\ref{sec:uniform}, for example.

To appreciate the meaning of the chain directions at the boundaries,
it may be helpful to consider the application of force in the
following way.  Imagine a packing of grains between to flat plates
that are perforated with holes much smaller than the grain size and
held at fixed separation.  Now imagine poking needles through the
holes in the plates, with the force applied to each needle being
specified externally and an equal total force applied from above and
below.  In this situation, there is a clear difference between chains
that end on needles and chains that end on a contact between a grain
and one of the plates.  The former must be present due to the boundary
conditions, and are therefore ingoing.  (See Figure \ref{fig:annihilation}.)
The latter
exist only as a response to the applied forces.  They would shift
around if a new needle were poked in initiating a new chain and hence
are not specifiable as boundary conditions .  Now in the case of two
rigid, smooth flat plates that are simply pushed together, it is not
clear how to distinguish the ingoing from the outgoing chains.
Nevertheless, for present purposes, we assume (plausibly, we think)
that such a distinction is somehow embedded in whatever physics at the
grain scale is responsible for the organization of stress into force
chains in the first place.  Exploration of the alternative hypothesis
-- that ingoing and outgoing chains are indistinguishable -- is beyond
the scope of this paper.  It would require self-consistent choices of
the splitting and fusion functions so as to yield structures that
satisfy the same Boltzmann equation when the directions of all force
chains are reversed.

\subsection{Rescalings and the importance of nonlinear terms}
\label{sec:rescaling}
In the isotropic \eqname equation, the parameter $\lambda$ can be set
equal to unity without loss of generality via a rescaling of lengths:
$x\rightarrow x/\lambda$, $z\rightarrow z/\lambda$, $P\rightarrow
\lambda P$ and $\mu \rightarrow \mu/\lambda$.  In these new units,
$\mu$ has dimensions of force.

A crucial feature of Eq.~(\ref{eqn:master}) is that the fixed point
solution $P(\fv)=0$ is unstable.  Consider the total force density in
the system, defined as ${\cal F}=\int\!df d\theta d\rv f P(\fv)$.  For
$P(\fv)<<1$, the value of ${\cal F}$ is determined by the linear terms
in the \eqname equation.  But these describe only splittings, and every
splitting increases ${\cal F}$ since force balance requires that the
sum of the two outgoing intensities be larger than the incoming
intensity.  If we begin with a single chain, each generation of
splittings increases ${\cal F}$ by a constant factor that depends upon
$\psi_{s}$ but is always greater than unity.  In a sufficiently large
sample, the rate of increase of ${\cal F}$ due to splitting will
exceed the rate of decrease due to leakage through the boundary.
In order to regulate this divergence, the nonlinear terms must be
included.  This argument will be made precise for the choice
of $\psi_{s}$ discussed in Section~\ref{sec:discrete} and developed 
in somewhat more general terms in the appendix.

The inclusion of the nonlinear terms requires the introduction of the
dimensionful parameter $\mu$.  To understand how this parameter is
determined, it is helpful to write it as the product of two factors,
$\mu=\mu_0 Y$, where $\mu_0$ has units of force and $Y$ is simply a
numerical constant.  Since there is no intrinsic force scale in the
system, $\mu_0$ can be chosen arbitrarily without loss of generality.
This freedom is a direct consequence of the fact that, given any DFCN,
there exists a continuous family related to it by multiplication of
all of the chain intensities by a constant factor. The rescaling associated
with the choice of $\mu_0$ is exactly compensated by a rescaling of the
argument $f$ in $P(f,\theta,\rv)$, so that the geometric structure of the
DFCN is unchanged.

The factor $Y$ is determined by the overall probability of fusions
occurring when chains intersect.  $Y$ can be thought of as a
normalization associated with $\phi_{f}$.  In physical terms, changes
in $Y$ have real effects on the $P$'s, as $Y$ ultimately determines
the density of fusions in the system.  It is convenient, however, to
work with variables $P^{\prime}\equiv P/Y$.  This results in the same
\eqname equation with $Y=1$, which we will work with henceforth while
dropping the primes on the $P$'s.  Thus we are left with a \eqname
equation with no parameters other than the choices of $\psi_s$ and
$\psi_f$: for the purposes of mathematical analysis $\lambda$ and
$\mu$ can both be scaled to unity.

Eq.~(\ref{eqn:master}), with $\lambda$ and $\mu$ both scaled to $1$,
but with generic forms of $\psi_{s}$ and $\psi_{f}$, has proven
difficult to solve for even the simplest sample geometries and
boundary conditions (though the authors still hold onto hope for
further analysis).  Moreover, the numerical generation of directed
force chain networks encounters serious difficulties for systems deep
enough that the nonlinear processes become important.  Useful insights
can be obtained, however, by considering some special cases where
analytical progress is possible.

\subsection{A solvable case: $120^{\circ}$ splittings and fusions}
\label{sec:discrete}
We now consider a system for which
\begin{eqnarray}
\psi_{s,f}(\theta_1\,|\,\theta_2,\theta_3) & = &
\frac{1}{2}\left[ 
\delta(\theta_{21}-\frac{\pi}{3})\delta(\theta_{31}+\frac{\pi}{3})\right. \nonumber \\
\ & \ & \;\; \left. + \delta(\theta_{21}+\frac{\pi}{3})\delta(\theta_{31}-\frac{\pi}{3})
\right]\,,
\end{eqnarray}
where $\theta_{ij}\equiv\theta_i-\theta_j$.  This corresponds to the case where
both splittings and fusions are always symmetric and form vertices
composed of $120^{\circ}$ angles, as in Fig.~\ref{fig:forcechains}.
Note that in this case, chains with different intensities or with
directions that are not parts of the single symmetric star of six
unit vectors, cannot interact via fusion.  So we further assume that
all chains initiated at the boundary have directions chosen from a
single ``6-fold star'' and have identical intensities $f_{\ast}$:
\begin{equation}
P(f,\theta) = \sum_{n=0}^{5} P_n\delta(f-f_{\ast})\delta(\theta-\theta_n),
\end{equation}
where $\theta_n\equiv (n-1/2)\pi/3$ as illustrated in
Fig.~\ref{fig:stars}(a).  We refer to this orientation of the
6-fold star vectors as ``horizontal'', since it includes directions
oriented horizontally.  Another choice we have studied extensively is
the ``vertical'' orientation shown in Fig.~\ref{fig:stars}(b).  It
turns out that for the horizontal orientation, exact analytical
expressions are easier to obtain, but the response functions (computed
numerically for the vertical orientation) are qualitatively similar.
\setlength{\smallfig}{0.0cm}
\addtolength{\smallfig}{0.8\linewidth}
\begin{figure}
\begin{center}
\epsfxsize=\smallfig
\epsfbox{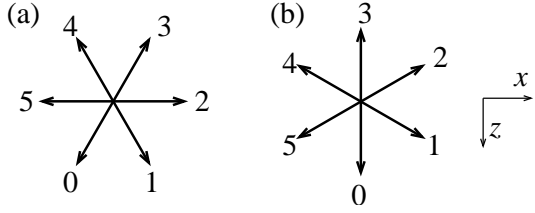}
\caption{Two choices of the discrete set of six permitted
directions of force chains.
\label{fig:stars}}
\end{center}
\end{figure}

For the horizontal orientation, Eq.~(\ref{eqn:master}) reduces to
\begin{eqnarray}
\vh_{i}\cdot\nabla P_{i} & = & -P_{i}+P_{i+1}+P_{i-1} \nonumber \\
\ & \ & + \left(P_{i+1}P_{i-1}-P_{i}P_{i+2}-P_{i}P_{i-2}\right),
\label{eqn:discretemaster}
\end{eqnarray}
where all indices are taken modulo 6 and $\vh_{i}\equiv
(\cos\theta_i,\sin\theta_i)$.  Note that all integrations over $f$ and $\theta$
have been performed, and the only remaining difference between the
coefficients of splitting and fusion events is the factor $Y$, which
has been absorbed into the $P$'s.

The restriction of the possible orientations of force chains to six
directions does {\em not} imply that the chains reside on a regular
geometric lattice.  The splitting and fusion events occur at arbitrary
positions, governed only by the probabilistic rules encoded in the
\eqname equation.  

\section{Horizontally uniform solutions}
\label{sec:uniform}
Let us now construct solutions of Eq.~(\ref{eqn:discretemaster}) for
the case of an infinite horizontal slab subject to a horizontally
uniform load.  Let $z=0$ represent the top surface of the slab and
$z=d$ the bottom.  For simplicity, we consider the case where the
loading (and hence the full solution) is symmetric under reflection
through a vertical line, as well as translationally invariant in the
horizontal direction.  Thus we have $P_0=P_1$, $P_5=P_2$, and
$P_4=P_3$ at all points in the system and $\partial_x P_n=0$ for all
$n$.  

The \eqname equation then reduces to the following set:
\begin{eqnarray}
\frac{\sqrt{3}}{2}\partial_z P_1 & = & P_2 - P_1 P_3\,, \label{eqn:h1} \\
 P_2 & = & P_1 + P_3 + \left[P_1 P_3 - P_2(P_1+P_3)\right]\,, \label{eqn:h2}\\
-\frac{\sqrt{3}}{2}\partial_z P_3 & = & P_2 - P_1 P_3\,, \label{eqn:h3}
\end{eqnarray}
where each of the $P$'s is a function of $z$ alone.  From
Eqs.~(\ref{eqn:h1}) and~(\ref{eqn:h3}), one sees immediately that
$G\equiv P_1+P_3$ is a constant (determined by the boundary
conditions), and from Eq.~(\ref{eqn:h2}) we have
\begin{equation}\label{eqn:hp2}
P_2 = \frac{G + P_1 P_3}{1 + G}\,. 
\end{equation}
Substituting into Eqs.~(\ref{eqn:h1}) and~(\ref{eqn:h3}) leads
to the two coupled ordinary differential equations for $P_1(z)$ and $P_3(z)$:
\begin{eqnarray}
\frac{dP_1}{dz} & = & \frac{2 G}{(1+G)\sqrt{3}}\left(1- P_1 P_3\right) \nonumber \\
\frac{dP_3}{dz} & = & \frac{-2 G}{(1+G)\sqrt{3}}\left(1- P_1 P_3\right)\,. \label{eqn:hp13}
\end{eqnarray}
These equations are to be supplemented with boundary conditions on
$P_1$ at $z=0$ and on $P_3$ at $z=d$.  ($P_1$ and $P_3$ are the
densities of downward and upward directed chains, respectively.)
From Eq.~(\ref{eqn:stresstensor}), the stress components are simply
\begin{eqnarray}\label{eqn:hhsigma}
\sigma_{zz} & = & \frac{3}{2}(P_1 + P_3)\,, \\
\sigma_{xx} & = & 2 P_2 + \frac{1}{2}(P_1 + P_3)\,,  \\
\sigma_{xz} & = & 0\,.
\end{eqnarray}

Before proceeding to the full solution of Eqs.~(\ref{eqn:hp13}), it is
instructive to consider the linearized theory in the vicinity of
$P_{i}=0$:  
\begin{equation}
\partial_z P_{1}=\frac{ 2}{\sqrt{3}}G;\quad 
\partial_z P_{3}=\frac{-2}{\sqrt{3}}G.
\end{equation}
Since $G\equiv P_1+P_3$ is a positive constant, one sees immediately
that $P_3$ will become negative for sufficiently large $z$.  In fact,
for arbitrarily large values of $P_3(0)$, one sees immediately that
$P_3(d)$ must become negative for $d>\sqrt{3}/2$, since the smallest
possible value of $G$ is $P_3(0)$.  Thus the linear theory produces
unphysical results for systems of dimensionful depth
$\lambda\sqrt{3}/{2}$ or greater.  The problem can be traced to the
divergence of chain densities mentioned above that occurs when the
system is sufficiently deep to allow an appreciable number of force
chains to ``turn around''.

In marked contrast to the linear theory, the full theory expressed in
Eq.~(\ref{eqn:hp13}) admits physical solutions for all possible
specifications of $P_1(0)$ and $P_3(d)$ for arbitrarily large $d$.
Let $B$ denote $P_1(0) P_3(d)$.  The solutions are
\begin{equation}\label{eqn:solns}
P_1 = \left\{ \begin{array}{ll}
\frac{G}{2} - \gamma_{+}\tanh\left[\frac{2}{\sqrt{3}}\frac{G \gamma_{+}}{(1+G)}(z-C)\right] & \! G > 2,\, B>1 \rule[-5mm]{0mm}{10mm}\\
\frac{G}{2} - \gamma_{+}\coth\left[\frac{2}{\sqrt{3}}\frac{G \gamma_{+}}{(1+G)}(z-C)\right] & \! G > 2,\, B<1 \rule[-5mm]{0mm}{10mm}\\
\frac{G}{2} + \gamma_{-}\tan \left[\frac{2}{\sqrt{3}}\frac{G \gamma_{-}}{(1+G)}(z-C)\right] & \! G < 2\,\end{array}\right.
\end{equation}
where $\gamma_{\pm}\equiv \sqrt{\pm (G^2 - 4)}/2$ and $G$ and $C$
are constants determined by the specification of $P_1(0)$ and
$P_3(d)$.  

We will see below that a solution with non-negative densities exists
for arbitrary choices of $P_1(0)$ and $P_3(d)$ and arbitrary values of
$d$.  On the other hand, for choices of $P_3(0)$ and/or $P_1(d)$
within some ranges, the equations lead to negative densities or no
solutions at all.  This is a mathematical expression of the physically
plausible statement that one cannot specify in advance the densities
of chains directed outward from boundary.

\begin{figure}
\begin{center}
\epsfxsize=\linewidth
\epsfbox{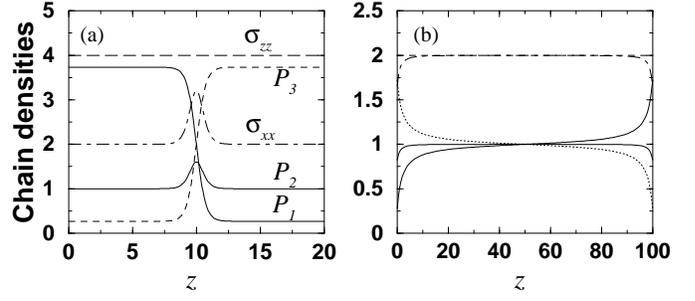}
\caption{Typical solutions of the discrete \eqname equation
  with symmetric boundary conditions.  (a) A generic solution with an
  exponentially localized transition region in the middle of the slab.
  (b) A solution with power-law boundary layers and a plateau in the
  middle, corresponding to $P_1+P_3=2-\epsilon$.
\label{fig:soln}}
\end{center}
\end{figure}

Exactly which specification of $P_1(0)$ and $P_3(d)$ correspond to a
given physical situation is not entirely clear.  For a slab squeezed
between two plates, however, a natural assumption is that the boundary
conditions should be symmetric: $P_3(d)=P_1(0)$.  In this case,
$C=d/2$.  Two particular solutions for symmetric boundary conditions
are shown in Fig.~\ref{fig:soln}.  Part (a) shows the case $G>2$,
which yields extended flat regions in the upper and lower halves of
the slab that are connected by a transition region.  The flat regions
correspond to a uniform stress that is anisotropic:
$\sigma_{xx}\neq\sigma_{zz}$.  The width of the transition region is
of order $\lambda$, with an exponentially fast approach to the
plateaus on either side.  Surprisingly, the transition region is
marked by a strong variation in the horizontal stress.  Though the
effect of such a transition region on the dynamical properties of the
system is far from clear, it is interesting to note that the theory
gives rise to an localized inhomogeneity that could influence shear
band formation. 
The increased horizontal stress in the transition region can be
intuitively understood by considering a case of very large $P_1(0)$
and $P_3(d)$.  In such a case, nearly all the chains in the top half
of the sample are downward pointing, and nearly all the chains in the
bottom half are upward pointing.  The density of horizontal chains is
determined by the splitting length, $\lambda$.  In the transition
region, however, there is a high density of intersections of upward
and downward pointing chains, which result in fusions that generate
horizontal chains.  Hence the density of horizontal chains is higher
in the transition region, being determined by the distances chains
propagate before fusing, rather than the longer distance required for
splitting.
Fig.~\ref{fig:soln}(b) shows density profiles for
$G<2$.  In this case, it is the edges that deviate from the fixed
point plateau in the middle, but the approach to the fixed point is a
power law, $\sim 1/z$, so one cannot clearly define a boundary layer.

The complete structure of the solution space is displayed in
Fig.~\ref{fig:traj}.  All points with $P_1 P_3 = 1$, shown as a
thick dotted line, are fixed points.  Since $P_1+P_3$ is a constant,
each trajectory is a line with slope $-1$.  The arrows indicate the
direction in which the systems moves with increasing $z$.  For
trajectories in the region above the fixed line, $P_1$ decreases with
depth and $P_3$ increases, and vice-versa for trajectories below (or
to the left of) the fixed line.  For small systems, trajectories in
the lower left corner, near the origin, may be relevant.  For deep
systems, however, the relevant trajectories must be ones that pass
very close to some fixed point where $P_1$ and $P_3$ can stay nearly
stationary for a long ``time''.
\setlength{\smallfig}{0.0cm}
\addtolength{\smallfig}{0.7\linewidth}
\begin{figure}
\begin{center}
\epsfxsize=\smallfig
\epsfbox{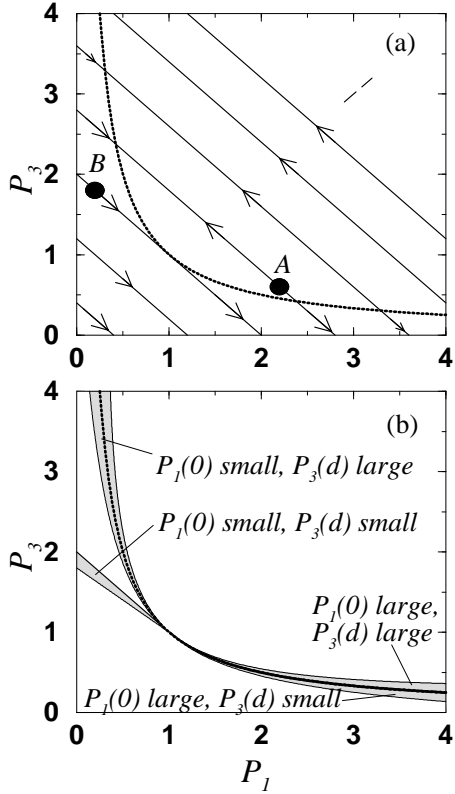}
\caption{(a) Trajectories of Eq.~(\ref{eqn:hp13}) in the $P_1$-$P_3$ plane.  
  The curved line, $P_1 P_3 = 1$ is a set of fixed points.  The
  diagonal lines indicate the trajectories, with the arrows showing
  the direction of flow with increasing $z$.  For systems deeper than
  $\lambda$, the typical length a force chain propagates before
  splitting, trajectories must either start very close to the line of
  fixed points (e.g. at point $B$), or on a trajectory that will take
  the system close to a fixed point (e.g. at point $A$).  (b) The
  shaded regions indicate where trajectories must start in order to
  satisfy various possible boundary conditions for deep systems.  The
  starting points must lie either in the vicinity of the fixed line,
  or just below the line of slope $-1$ tangent to the fixed line.  For
  $P_1(0)$ small and $P_3(d)$ large, the starting point may be on
  either side of the fixed line.
\label{fig:traj}}
\end{center}
\end{figure}

Let us now consider the possibilities for satisfying various boundary
conditions.  Let $U$ be the ``starting'' point $(P_1(0),P_3(0))$ and
$V$ be the ``ending'' point $(P_1(d),P_3(d))$.  First, we show that
for any $d$, any $P_1(0)$, and any $P_3(d)$, $U$ can be chosen in the
first quadrant in such a way that $V$ lies in the first quadrant;
i.e., there exists a trajectory with no negative chain densities that
satisfies the boundary conditions.  To see this, first note that
$P_1(d)$ can never be negative.  If $U$ is chosen below the fixed
line, $P_1$ always increases with increasing $z$.  On the other hand,
since trajectories can never cross the fixed line, if $U$ is above the
fixed line, the entire trajectory is confined to the first quadrant.
To complete the proof, it is sufficient to show that as $U$ varies in
the first quadrant along the vertical line corresponding to any given
choice of $P_1(0)$, $P_3(d)$ takes on all possible positive values.
Consider $U=(x,y)$.  For any nonzero value of $d$, $P_3(d)$ is
negative for $y=0$, since all trajectories beginning on the $x$-axis
pass immediately into the fourth quadrant.  As $y$ is increased toward
the fixed point $y=1/x$, $P_3(d)$ rises continuously to $1/x$.  As $y$
is increased beyond $1/x$, $P_3(d)$ increases without bound.  Hence,
given $P_1(0)=x$ one can always choose $y$ such that $P_3(d)$ has any
specified positive value.  Fig.~\ref{fig:traj}(b) indicates roughly
where $U$ must be chosen in order to satisfy various boundary
conditions when $d$ is large.  For $d$ small or of the order of unity,
the trajectories generally lie completely within the unshaded regions
of the diagram.

Suppose now that we attempt to specify $P_1$ and $P_3$ at the top
boundary.  In other words, we specify $U$ completely.  In this case,
problems arise for sufficiently large $d$ whenever $U$ is below the
fixed line for $P_1(0)>1$ or below the line $P_1+P_3=2$ for
$P_1(0)<1$.  Each trajectory in these regions passes into the fourth
quadrant at some finite value of $z$, generating the unphysical result
that $P_3$ becomes negative.  Similar difficulties arise if one
attempts to specify $V$ rather than $U$.  Finally, if one attempts to
specify $P_1(d)$ and $P_3(0)$, there are no solutions at all for $d$
large and $P_1(d)>P_3(0)+1/P_3(0)$.  The fact that certain regimes are
prohibited in each of these cases is consistent with the assertion
that physical constraints permit the specification of $P_1(0)$ and
$P_3(d)$ only.

For deep systems (large $d$), there are three types of typical
solutions, corresponding to the three expressions in
Eq.~(\ref{eqn:solns}):
\begin{description}
\item[Type I:] $P_1(0)> 1$ and $P_1(0)P_3(d)>1$.  The motion
  must take place in the region above the line of fixed points,
  starting close to the line, at a point such as $A$ in
  Fig.~\ref{fig:traj}.  For symmetric boundary conditions, the
  motion will also end near a fixed point, producing the behavior
  shown in Fig.~\ref{fig:soln}(a).
\item[Type II:] $P_1(0)> 1$ and $P_1(0)P_3(d)<1$.  The
  motion takes place in the lower right portion of the plane.  The
  trajectory must start just below a fixed point.  Such solutions
  occur if the bottom boundary condition is taken to be $P_3(d)=0$.
  Following such a trajectory backwards (up from the bottom of the
  sample) one finds the densities approach their fixed point values
  like $1/z$.  We will not pursue this case further here, but note
  that related asymmetric solutions might be relevant when gravity is
  included in the model.
\item[Type III:] $P_1(0)<1$ and $P_3(d)<1$.  For large $d$, the motion
  must take place on a trajectory just below the one marked $B$ in
  Fig.~\ref{fig:traj}, so as to pass very close to the fixed point
  at $P_1=P_3=1$.  The density profiles associated with these
  trajectories are similar to those of Type II, but show increasing
  deviations from the fixed point values at both the top and bottom of
  the slab.  Density profiles for a Type III solution are shown in
  Fig.~\ref{fig:soln}(b).  (Note the difference in horizontal scale
  between parts (a) and (b).)  Though these solutions are
  mathematically consistent, they have the counter-intuitive physical
  property that the density of upward-pointing chains at the top
  surface is larger than the imposed density of downward-pointing
  chains.
\end{description}
All other solutions are related to these three types by reflection through
$z=d/2$ and interchange of $P_1$ and $P_3$.

For deep systems, the system is primarily composed of regions in which
$(P_1,P_3)$ lies close to a fixed point.  The nature of the directed
force chain networks at the different fixed points is therefore of
interest.  First, we note that the networks associated with generic
fixed points are anisotropic, having different densities for chains in
different directions.  Letting $p$ designate the value of $P_1$ at a
fixed point, the value of $P_3$ at that fixed point is $1/p$.  We
emphasize that this is not a result of intrinsic anisotropy in the
material, but rather a result of the boundary conditions.  The
situation is similar to that of an elastic material that may be
isotropic when unloaded, but becomes anisotropic when subjected to a
uniaxial stress.  There is a crucial difference, however, in that the
force chain network has no stable unloaded state.  Thus, except under
conditions of pure hydrostatic pressure, the response of the force
chain network to perturbations in its loading will generally be
anisotropic.  The exceptional case is the fixed point at $p=1$, where
the system is indeed isotropic.

Second, we may compute two distinct quantities that might be thought
of as analogous to the Poisson ratio of standard elasticity theory.
The first is the ratio $\Delta\sigma_{xx}/\Delta\sigma_{zz}$ obtained
upon increasing the strengths of the forces in all of the force
chains.  Since this does not change the network at all, the resulting
ratio is just equal to $\sigma_{xx}/\sigma_{zz}$ at the fixed point.
Using Eq.~(\ref{eqn:hhsigma}), we find
\begin{equation}
\nu_1 \equiv \frac{\sigma_{xx}}{\sigma_{zz}} = \frac{4+p+p^{-1}}{3(p+p^{-1})},
\end{equation}
which varies monotonically from $1$ at $p=1$ to $1/3$ for
$p\rightarrow\infty$.  

The other quantity analogous to the Poisson ratio is obtained by
considering small increases in the applied {\em density} of force
chains, rather than their strengths.  In this case, the system moves
along the line of fixed points and we calculate
\begin{equation}
\nu_2\equiv\frac{\partial_p\sigma_{xx}}{\partial_p\sigma_{zz}} = 1/3.
\end{equation}
It is amusing to note that a Poisson ratio of 1/3 arises also in
studies of ball-and-spring networks with energies expressible as a sum
of two-body central forces.\cite{limat} We shall see in
Section~\ref{sec:v6}, however, that this feature is specific to the
horizontal orientation of the 6-fold star vectors.  For the
vertical orientation, $\nu_2$ varies with $p$.

It is also interesting to note that $\nu_1$ and $\nu_2$ lie within the
range of stability for 2D isotropic elasticity, in contrast to the
Poisson ratio computed in Ref.~\cite{bcol} from the linear theory of
DFCN's.  This is an indication that there are significant differences
between the elasticity theories in the vicinity of a nontrivial fixed
point and that obtained in the vicinity of the origin.

\section{Response to a localized force}
\label{sec:response}
The response of the directed force chain network to a localized force
applied at the top boundary may be computed via linearization about
the appropriate fixed point.  In general, for a DFCN with a continuum
of force chain intensities, there are two distinct ways to apply a
localized perturbation: (1) by changing the strength of some of the
force chains injected at the boundary; or (2) by adding some new force
chains of chosen intensities.
In the case where splitting and fusion angles are always
$120^{\circ}$, however, option (1) is not available.
Because $\psi_s$ and $\psi_f$ permit interactions
only among force chains of the same intensity, increasing the strength
of one chain would be equivalent to removing that chain from the
existing network and creating an entirely new network completely
decoupled from the original, which would lead right back to
the problem of the failure of the linear theory in the vicinity of the
origin.  Thus we must consider option (2), in which we inject a low
density of new force chains with intensity $f_{\ast}$ along one or
more of the directions $\vh_i$.  

In the fixed point DFCN, the chain densities are translationally
invariant.  This permits the derivation of decoupled equations for the
different Fourier modes of a perturbation applied at the top surface.
We begin from the six equations obtained from
Eq.~(\ref{eqn:discretemaster}):
\begin{eqnarray}\label{eqn:hmaster}
\frac{\sqrt{3}}{2}\partial_z P_1 + \frac{1}{2}\partial_x P_1 & = & -P_1 + P_2 + P_0 \nonumber \\
\ & \ & + \left(P_0 P_2 - P_1 P_3 - P_1 P_5 \right),  \\
\partial_x P_2 & = & -P_2 + P_3 + P_1 \nonumber \\
\ & \ & + \left(P_1 P_3 - P_2 P_4 - P_2 P_0 \right), \\
\frac{-\sqrt{3}}{2}\partial_z P_3 + \frac{1}{2}\partial_x P_3 & = & -P_3 + P_2 + P_4 \nonumber \\
\ & \ & + \left(P_2 P_4 - P_3 P_5 - P_3 P_1 \right), 
\end{eqnarray}
and similar equations for $P_4$, $P_5$, and $P_0$.  Fixed point
solutions of these equations have the form
$(P_1^0,P_3^0,P_4^0,P_0^0)=(p,1/p,1/p,p)$, with $P_2^0=P_5^0$ given by
Eq.~(\ref{eqn:hp2}).  Let $p_n = P_n(x,z)-P_n^0(x,z)$ be the
deviations from the fixed point and define Fourier coefficients
$Q_n(q,z)$ via
\begin{equation}
p_n = \int_{-\infty}^{\infty}\!\!dq\; Q_n(q,z)\; e^{iqx}\,.
\end{equation}
Linearization of Eq.~(\ref{eqn:hmaster}) in $Q$ yields 
\begin{equation}\label{eqn:linearh}
\frac{d}{dz}\Qv(q) = \Mm(q)\cdot\Qv(q),
\end{equation}
where $\Qv$ is the column vector $(Q_1, Q_3, Q_4, Q_0)$ and the
complex elements of $\Mm$ are complicated algebraic functions of $p$
and $q$.  Analytic expressions for the eigenvalues of $\Mm$ may be
obtained, but they are exceedingly complicated functions of $p$ and
$q$ and we do not reproduce them here.  (They involve solutions of
quartic equations whose coefficients are polynomials of seventh order
in $p$ and third order in $q$.)  The eigenvalues and eigenvectors of
$\Mm(q)$ and $\Mm(-q)$ are related by the condition that the response
to a real perturbation must be real.

To compute the response to a localized perturbation applied at the top
of a semi-infinite system, we seek solutions for which all $Q_n$'s
vanish as $z\rightarrow\infty$.  The symmetry of the \eqname equation
and the fixed point solution under $x\rightarrow -x$ guarantees that
for every complex eigenvalue $\kappa$ of $\Mm(q)$, there will be a
complex conjugate partner $\kappa^{\ast}$.  In addition, though the
$z\rightarrow -z$ symmetry is broken by the fixed point solution
(since $P_1^0\neq P_3^0$), there are two eigenvalues with negative
real part for each $q$ and two with positive real part.  In
constructing the perturbation, the coefficients of all eigenvectors
having eigenvalues with positive real parts must be zero, since for
these modes the $Q_n$'s grow exponentially with increasing $z$.  By
taking appropriate linear combinations of the eigenvectors associated
with the eigenvalues that have negative real parts, one can arrange to
satisfy any specified boundary condition on $p_1(x,0)$ and $p_0(x,0)$.
Note that $p_3(x,0)$ and $p_4(x,0)$ are then determined, which is
consistent with the general rule that we may specify the densities of
inward-directed chains only.

For simplicity, we restrict the presentation here to applied
perturbations that are symmetric under reflection about a vertical
line: perturbations with $p_0(x) = p_1(-x)$.  The general form
of the solution for $\pv = (p_1,p_3,p_4,p_0)$ is then
\begin{equation}\label{eqn:responseform}
\pv(x,z) = \int_{0}^{\infty}\!\!dq\;\sum_{j=1}^{2}
a_j(q)\, e^{{\rm Re}[\kappa_j]z}\cos(qx - {\rm Im}[\kappa_j]z)\, \ev_j, 
\end{equation} 
where $\kappa_j$ is an eigenvalue with negative real part, $\ev_j$ is
its associated eigenvector, and $a_j(q)$ is a constant determined by
the boundary conditions.

\begin{figure}
\begin{center}
\epsfxsize=\linewidth
\epsfbox{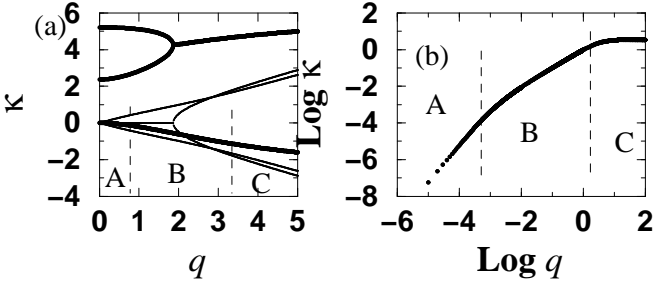}
\caption{The spectrum of eigenvalues of the linear operator governing
  perturbations in the vicinity of a fixed point for the case of the
  horizontal orientation of 6-fold star vectors.  (a) The case
  $P_1=3$.  Thick (thin) lines are the real (imaginary) parts of
  $\kappa$.  A, B, and C mark regimes corresponding to different
  behavior of the response function.  (See text.)  (b) The case
  $P_1=1.001$.  Only the real part of $\kappa$ is shown.  The
  quadratic, linear, and flat behaviors in the regimes A, B, and C are
  clearly visible.
\label{fig:hhspectrum}}
\end{center}
\end{figure}

Fig.~\ref{fig:hhspectrum}(a) shows the eigenvalues $\kappa (q)$ of
$\Mm$, computed for the case $p=3$.  The thick lines indicate the real
parts of $\kappa$ and the thin lines the imaginary parts.  The lower
thick line and the two thin lines that emerge from the origin above and
below the axis correspond to a complex conjugate pair of roots, the
two roots with negative parts and therefore the only ones relevant for
the response in a semi-infinite system.  The spectrum may be divided,
roughly, into three regions.  In region A, corresponding to the
smallest values of both $q$ and $|{\rm Re}[\kappa]|$, neglecting all
terms of order $q^3$ or higher in $\Mm$ we find
\begin{equation} \label{eqn:kab}
\kappa\approx -D q^2 + i c q ,
\end{equation} 
where $c$ and $D$ are real constants:  
\begin{eqnarray}
D  & = & \frac{4}{\sqrt{3}}\frac{1}{(p^2-p^{-2}}; \nonumber \\
c  & = & \frac{1}{\sqrt{3}}. \label{eqn:hab}
\end{eqnarray}
There is a transition region $B$ with intermediate
$q$ and $|{\rm Re}[\kappa]|$, and finally, region C, a plateau in
$|{\rm Re}[\kappa]|$ for large $q$.

The response at large $z$ is dominated by the slowest decaying modes,
those with $|{\rm Re}[\kappa]|$ closest to zero.  From the observation
that there are no values of $q$ other than zero for which ${\rm
  Re}[\kappa]$ vanishes, it is clear that these are the long
wavelength modes with the disperion relation of Eq.~(\ref{eqn:kab}).
Thus for large $z$ we can approximate the integral in
Eq.~(\ref{eqn:responseform}) by
\begin{eqnarray}\label{eqn:responselargez}
\pv & \sim & \int_{0}^{\infty}\!dq\; 
e^{-D z q^2}
\left[\cos(q(x - c z)) \ev_1 \right. \nonumber \\ 
\ & \ & \quad\quad\quad\quad\quad\quad\quad\quad\quad\quad\left. + \cos(q(x - c z)) \ev_2\right] \\
 & \sim & z^{-1/2}\left[e^{-\frac{(x- c z)^2}{4 D z}}\ev_1
 + e^{-\frac{(x+c z)^2}{4 D z}}\ev_2\right]. 
\end{eqnarray}
The response at large depths consists of two Gaussian peaks that
propagate away from $x=0$ at an angle of $\tan^{-1}c$ and have
widths proportional to $\sqrt{D z}$.  

Eq.~(\ref{eqn:hab}) indicates that $c$ corresponds to angles of
propagation along the directions $\vh_1$ and $\vh_0$ for all of the
fixed points.  This appears to be special to the horizontal
orientation of the star vectors.  (See Section~\ref{sec:v6} below.)
$D$ diverges as the isotropic fixed point, $p=1$, is approached.
This means that for nearly isotropic fixed points, the peaks at large
$z$ become increasingly broad.  In addition, the region over which the
spectrum is quadratic (region A), becomes smaller and smaller, so that
the emergence of the two peaks occurs only for exceedingly large $z$.
Fig.~\ref{fig:hhspectrum}(b) shows the real part of the spectrum for
the case $p=1.001$, plotted on a logarithmic scale.  In this case we
see that the transition region B is a clearly defined regime of linear
dispersion.  Linear dispersion generally implies either a single,
centered peak in the response, or propagating peaks that broaden
linearly with depth.

The full response to a localized applied force can be computed
numerically.  Fig.~\ref{fig:hhresponse} shows the response function
for the case $p=3$.  The curves were obtained by approximating the
integral over $q$ in Eq.~(\ref{eqn:responseform}) with a sum over
discrete values $q=(n-1/2)\delta q$, with $\delta q=0.01$ and $1\leq
n\leq 1000$.  This range of $q$'s contains enough high $q$'s to
construct a relatively sharp initial Gaussian at $z=0$, and enough low
$q$'s to observe the response at large $z$.  For each $q$, the
eigenvalues and eigenvectors of $\Mm$ are determined.  We then obtain
the linear combination of the two stable eigenvectors required to
produce the vector $\Qv=(1,y,y,1)$, where $y$ is unspecified.  (The
two conditions $p_1(q)=1$ and $p_0(q)=1$ determine the linear
combination of the two eigenvectors completely, and hence determine
$p_3=p_4=y$, and $p_2$ and $p_5$ in turn.)  These linear combinations
of eigenvectors are then multiplied by a factor ($a_j=\exp(-q^2/10)$)
so that the sum in Eq.~(\ref{eqn:responseform}) at $z=0$ yields
centered Gaussians for $p_1$ and $p_0$, which constitute the applied
perturbation.  The response is then determined at various values of
$z$ by summing the integrand of Eq.~(\ref{eqn:responseform}) over the
discrete set of $q$'s.
\setlength{\smallfig}{0.0cm}
\addtolength{\smallfig}{0.7\linewidth}
\begin{figure}
\begin{center}
\epsfxsize=\smallfig
\epsfbox{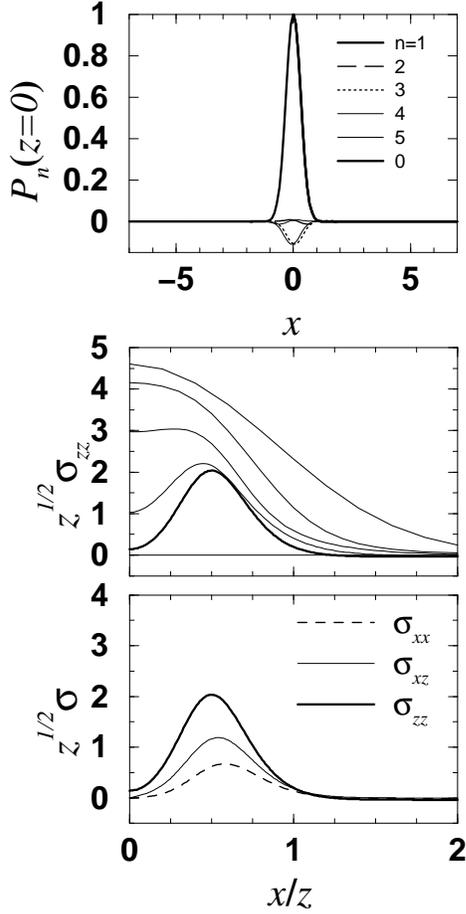}
\caption{Response functions for 6-fold star vectors in horizontal
  orientation.  (a) $P_n$'s at the top surface.  $P_1$ and $P_0$ are
  identical and are specified as boundary conditions.  The other
  $P_n$'s are part of the response.  Due to the symmetry of the
  boundary conditions, $P_4(x)=P_3(-x)$ and $P_5(x)=P_2(-x)$.  (b)
  Scaled profiles of $\sigma_{zz}$ for several values of $z$.  From
  top to bottom, the curves correspond to $z=0.2,0.5,1,2,5$, and $10$.
  (c) Different components of the stress tensor at $z=10$,
  corresponding to the thick line (b).  Only positive $x$ is shown.
  The symmetry of the boundary conditions implies
  $\sigma_{zz}(x)=\sigma_{zz}(-x)$, $\sigma_{xx}(x)=\sigma_{xx}(-x)$,
  and $\sigma_{xz}(x)=-\sigma_{xz}(-x)$.
\label{fig:hhresponse}}
\end{center}
\end{figure}

Fig.~\ref{fig:hhresponse}(a) shows the $p_n$'s at the top surface.
The identical, centered Gaussian $p_1$ and $p_0$ represent the
imposed boundary condition.  The other $p_n$'s at the top surface are
part of the response, as determined by the condition that $p_3$ and
$p_4$ must vanish at $z=\infty$.  Part (b) shows profiles of
$\sigma_{zz}$ along horizontal slices at various values of $z$.  The
curves have been scaled so that the trend toward diffusively
broadening, linearly propagating peaks is evident.  At large $z$, the
peaks in this figure would become increasingly sharp while maintaining
their same amplitude and position.  Part (c) shows the relation of the
various stress components at large $z$.

\section{Other choices of star vectors}
\subsection{6-Fold star in the vertical orientation}\label{sec:v6}
After the horizontal orientation of the 6-fold set of star vectors,
the next simplest case is the vertical orientation of the 6-fold star.
Calculations for the vertical orientation are an important means of
testing the robustness of several features of the solutions found
above.  For horizontally uniform systems with reflection symmetry
under $x\rightarrow -x$, the \eqname equation reduces to a set of four
coupled, nonlinear, ordinary differential equations for the variables
$P_0$, $P_1$, $P_2$, and $P_3$.  (See Fig.~\ref{fig:stars}.)  Using
the fact that $\sigma_{zz}$ is conserved, these can be reduced to a set
of three. Exact analytical solution for the chain densities as a function of
depth for generic boundary conditions is difficult.  Nevertheless, it
is easy to find a line of fixed points, which is all that is required
for carrying out an analysis of the response as performed above for
the case of the horizontal 6-fold star.  The fixed points identified
below were found by making the assumption that the densities of chains
in opposite directions are related by multiplicative inversion; i.e.,
$P_0 P_3 = P_1 P_4 = P_2 P_5 = 1$.  (This was inspired by the
horizontal case, in which exactly analogous relations were found to
hold, but we have not ruled out the possibility that other fixed
points exist.)

\begin{figure}
\begin{center}
\epsfxsize=\linewidth
\epsfbox{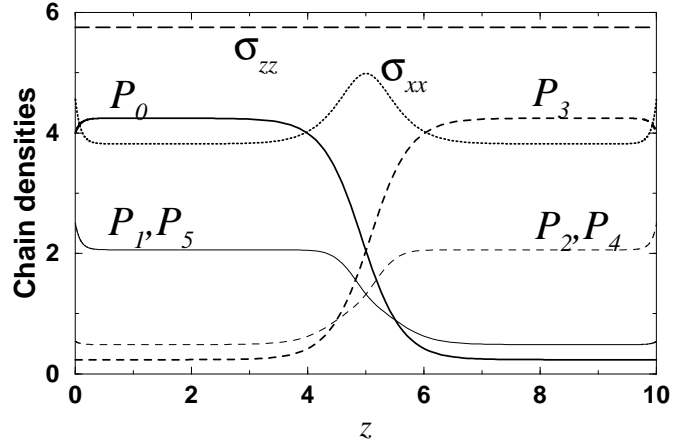}
\caption{A typical solution of the discrete \eqname equation
  with symmetric boundary conditions for the vertical orientation of
  6-fold star vectors.  (Compare to Fig.~\ref{fig:soln}).  The
  boundary conditions are $P_0(0)=4$, $P_1(0)=P_5(0)=2.5$, $P_3(d)=4$,
  $P_2(d)=P_4(d)=2.5$, with $d=10$.  The transition region in the
  middle creates a bump in $\sigma_{xx}$.  Boundary layers are also
  present.
\label{fig:solnv}}
\end{center}
\end{figure}

Numerical solution of the two-point boundary value problem yields
solutions of the type shown in Fig.~\ref{fig:solnv}.  The basic
features of the solution are quite similar to those for the horizontal
orientation of the star vectors.  In particular, there are flat
regions corresponding to fixed point solutions and a transition region
in the middle of the slab with enhanced horizontal stress.  In the
vertical case, however, boundary layers at the top and bottom appear
as well.  These arise because the additional dimensions in $P$-space
allow for trajectories that begin away from the fixed line, approach
it rapidly, then eventually diverge from it again; i.e., the fixed
line has both attracting and repelling directions.  The precise forms
of the chain density profiles in the boundary layers and in the
transition region are sensitive to the details of the imposed boundary
conditions, but the enhancement of $\sigma_{xx}$ in the transition
region is a robust feature.

The fixed points of Eq.~(\ref{eqn:discretemaster}) with $x\rightarrow
-x$ symmetry have
\begin{equation}
P_0 = p^2;\,\,P_1=p;\,\,P_2=p^{-1};\,\,P_3=p^{-2},
\end{equation}
which yield
\begin{eqnarray}
\nu_1 & = & \frac{3(p+p^{-1})}{2(p^2+p^{-2})+(p+p^{-1})}, \\
\nu_2 & = & \frac{3}{1+4(p+p^{-1})}.
\end{eqnarray}
for the two types of ``Poisson ratio''.  Here, as for the case of the
horizontal orientation of the 6-fold star vectors, $\nu_1$ varies
monotonically from unity to 0 as $p$ ranges from 1 (the isotropic
fixed point) to $\infty$.  Unlike the horizontal case, however,
$\nu_2$ is not constant.  It is again $1/3$ at the isotropic fixed
point, but decreases monotonically to zero for large $p$.

\begin{figure}
\begin{center}
\epsfxsize=\linewidth
\epsfbox{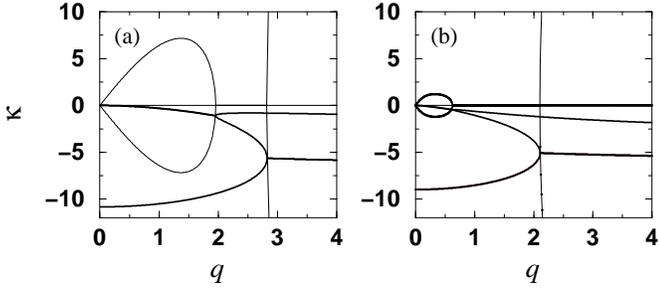} 
\caption{The spectrum of eigenvalues of the linear operator governing
  perturbations in the vicinity of a fixed point for the case of the
  vertical orientation of 6-fold star vectors.  Only the
  eigenvalues with negative real parts are shown.  (a) The case
  $P_1=2$.  Thick (thin) lines are the real (imaginary) parts of
  $\kappa$.  (b) The case $P_1=1.1$.  In both (a) and (b), the
  magnitudes of the imaginary parts have been multiplied by 15 for
  clarity.
\label{fig:hvspectrum}} 
\end{center}
\end{figure}

To determine the response function, we work in the vicinity of a
generic fixed point, neglecting the effects produced by the boundary
layer.  Fourier analysis of the linearized equations yields the
spectrum shown in Fig.~\ref{fig:hvspectrum}.  The spectrum at low
$q$ has the form of Eq.~(\ref{eqn:kab}), with
\begin{eqnarray}
D  & = & \frac{6(2r_{4+} +r_{3+} +8r_{2+} +9r_{1+}+14)}
                   {\left(4r_{1+} +1\right)^2\,\left(2r_{3-} +r_{2-}-2r_{1-}\right)},\nonumber \\
c  & = & \sqrt{3(4r_{1+}+1)^{-1}}. \label{eqn:vab}
\end{eqnarray}
where we have defined $r_{n\pm}\equiv p^n \pm p^{-n}$.  For strongly
anisotropic fixed points (i.e., large $p$) we have $c\approx
(3/8)p^{-1}$ and $D\approx (4p/3)^{-1/2}$, implying, at large $z$,
slowly diffusing peaks propagating close to the vertical direction.
For nearly isotropic fixed points $p=1+\epsilon$, we have $c\approx
\sqrt{1/3}\;(1-2\epsilon^2 /9)$ and $D\approx (1/3)\epsilon^{-1}$,
implying rapidly broadening peaks propagating along directions close
to halfway between the star vectors.  Thus, in constrast to the case
of the horizontal star orientation, the direction of propagation of
the peaks varies continuously from $30^{\circ}$ near the isotropic
fixed point to near $0^{\circ}$ at the strongly anisotropic fixed
points: the direction of propagation of the peaks is {\em not} tied to
the star vector directions.  

As in the case of the horizontal orientation, the region in $q$ space
over which the dispersion relation is quadratic shrinks to zero as the
isotropic fixed point is approached.  We also note that near the
isotropic fixed point, where $D$ becomes large, we can eliminate
$\epsilon$ to obtain $c\approx\sqrt{1/3}\;(1-(2/81)D^{-2})$, whereas
for the strongly anisotropic fixed points, where $D$ is small, we can
eliminate $p$ to obtain $c\approx D^2/2$.  These relations between
propagation direction and diffusion rate differ markedly from
relations obtain from models in which the characteristic propagation
direction and diffusion rates are both controlled by a parameter
corresponding to the strength of the disorder, where one finds
$c\approx c_0 - a D^{\alpha}$ with $\alpha > 0$. \cite{bcc,bccz} This
suggests that the DFCN is {\em not} in the weak disorder regime.

\begin{figure}
\begin{center}
\epsfxsize=\linewidth
\epsfbox{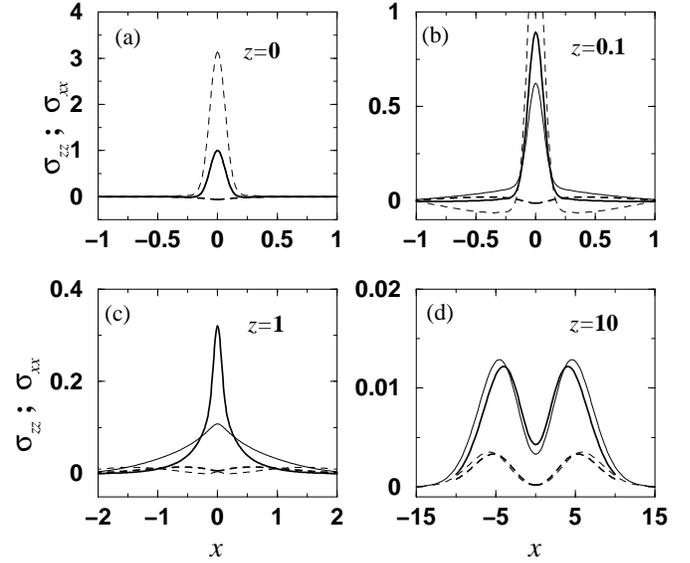}
\caption{Response functions for 6-fold star vectors in vertical
  orientation.  Solid lines indicate $\sigma_{zz}$ and dashed lines
  $\sigma_{xx}$.  Thick lines correspond to a boundary condition in
  which $P_0$ is a Gaussian of unit strength and $P_1$ and $P_5$ are
  $0$.  Thin lines correspond to a boundary condition in which $P_0=0$
  and $P_1$ and $P_5$ are Gaussian of strength 2, so as to make
  $\sigma_{zz}(z=0)$ the same for both cases.  Note the difference in
  both horizontal and vertical scales in the different panels.
\label{fig:hvresponse}}
\end{center}
\end{figure}

Typical response functions are shown in Fig.~\ref{fig:hvresponse}.
Two independent choices for the application of a localized vertical
load are shown:
\begin{equation}\label{eqn:vpertchoices}
(p_0,p_1,p_5) = \left\{ \begin{array}{c}
(\epsilon,0,0)\exp\left(-(x/0.09)^2\right) \\
{\rm and} \\
(0,2\epsilon,2\epsilon)\exp\left(-(x/0.09)^2\right)
\end{array}\right. .
\end{equation}  
These two choices
produce the same profile for $\sigma_{zz}(0)$, but different profiles
for $\sigma_{xx}(0)$, as shown in (a).  At the small depth shown in
(b), the two give different responses with somewhat complicated
structures.  At the intermediate depth shown in (c), the responses
differ as well, but both have something like the Lorentzian structure
familiar from the response of semi-infinite, standard elastic
materials.\cite{serero} In particular, both show a central peak with
long tails.  It is not until we reach depths significantly larger than
$\lambda$ (which has been scaled to unity), as in (d), that we see the
two diffusively broadening Gaussian emerge and the responses become
quite similar.

In the asymptotic regime at large depths, Fig.~\ref{fig:hvresponse}
suggests that $\sigma_{xx}$ becomes proportional to $\sigma_{zz}$.  In
fact, we can show both that $\sigma_{xx}\approx c^2 \sigma_{zz}$ and
that $c = \sqrt{\nu_2}$ by considering the eigenvectors at low $q$
associated with the dominant branch of the spectrum.  Let
\begin{eqnarray}
\sigma_{xx}(q) & = & \Sv^{(2)}\cdot\ev(q), \nonumber \\
\sigma_{zz}(q) & = & \Cv^{(2)}\cdot\ev(q); 
\end{eqnarray}
where $\Cv^{(2)}$ and $\Sv^{(2)}$ are row vectors with
components $\cos^2(\pi j/3)$ and $\sin^2(\pi j/3)$, respectively, and
$\ev(q)$ is the eigenvector of $\Mm(q)$ associated with the
dominant branch.  $\ev(q)$ converges as $q$ approaches zero
precisely to the $q=0$ mode corresponding to shifts along the fixed
line.  But these shifts are just the ones used to compute $\nu_2$.
Thus as $z$ increases and the surviving modes have $q$'s closer to
zero, and the ratio $\sigma_{xx}(q)/\sigma_{zz}(q)$ for all of those
modes approaches $\nu_2$.  To see that the ratio must equal $c^2$,
note that on large length scales the response appears as two narrow
lines that propagate along the directions $x=\pm cz$.  Newton's laws
require that the stress along these lines correspond to forces
directed along the line, which immediately implies
$\sigma_{xx}=c^2\sigma_{zz}$ along the line.
 
The crossover from a roughly standard elastic form to a hyperbolic one
with diffusive broadening is an unusual feature.  We note that linear
elasticity theories, even in the unstable regime where the equations
are hyperbolic, always yield scale invariant response functions for a
semi-infinite medium.  The spectrum is always purely linear, so that
the response is a function of $x/z$ only.  In the DFCN, however, the
spectrum has a richer structure and the response at large $z$ may
appear quite different from that at small or modest depths.  Though
the discrete model studied above does not represent a realistic
picture of a disordered granular material, the result does suggest
that experimental data should be interpreted cautiously: the
appearance of a response with a typical elastic form may be a
deceiving effect of working with a relatively small sample.

\begin{figure}
\begin{center}
\epsfxsize=\linewidth
\epsfbox{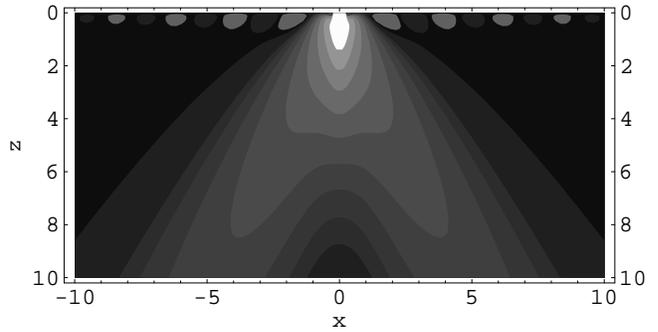}
\caption{Contour plot of $\sigma_{zz}$ for the vertical orientation of 
  6-fold star vectors an applied chain density
  $(p_0,p_1,p_5)=(0,2\epsilon,2\epsilon)\exp(-(x/0.09)^2)$.  The
  contour lines are at the set of values (-0.001, 0., 0.001, 0.005,
  0.007, 0.01, 0.014, 0.02, 0.028, 0.04, 0.056, 0.08).  The emergence
  of the double-peaked structure from a single peak between $z=0.5$
  and $4.0$ is clearly visible.  For $z<0.5$, a number of small,
  rapidly decaying peaks appear corresponding to wavenumbers near the
  broad peak in the spectrum near $q=2.5$, which is visible in
  Fig.~\ref{fig:hvspectrum}.
\label{fig:hvcontour}}
\end{center}
\end{figure}

Fig.~\ref{fig:hvcontour} shows a contour plot of the response for
the second choice of perturbation in Eq.~(\ref{eqn:vpertchoices}).
The figure clearly shows the emergence of the double-peaked structure
at large $z$ from a single peak at intermediate $z$.  It also shows
multiple peaks near $z=0$ that have not yet been mentioned.  These
peaks, which are of very low amplitude, form and decay rapidly.  The
feature in the spectrum that is responsible for them is the broad peak
in the uppermost branch of the real part in the neighborhood of
$q=2.5$.  (See Fig.~\ref{fig:hvspectrum}.)  Modes near this value of
$q$ decay slightly more slowly than others nearby.

\subsection{$n$-fold symmetric stars in 2D}
For sets of star vectors with the symmetry of a regular $n$-gon, as
depicted in Fig.~\ref{fig:nstar}, the full nonlinear problem is
much more difficult due to the fact that splitting generates chains
with different intensities.  Nevertheless, analysis of the \eqname
equation with fusion terms neglected can be carried out and yields
useful insights.
\setlength{\smallfig}{0.0cm}
\addtolength{\smallfig}{0.4\linewidth}
\begin{figure}
\begin{center}
\epsfxsize=\smallfig
\epsfbox{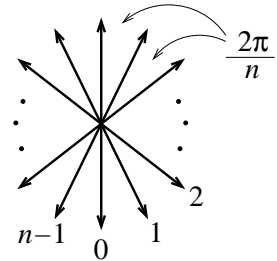}
\caption{The $n$-fold star used for analysis of the linear theory.  Each angle
is $2\pi/n$, where $n$ is an integer not divisible by 4.
\label{fig:nstar}}
\end{center}
\end{figure}

When the fusion terms in Eq.~(\ref{eqn:master}) are neglected, the
equation can be multiplied by $f$ and integrated over $f$ to obtain a
new \eqname equation governing the spatial variations in the quantity
$F(\theta)\equiv\int_0^{\infty}\!\! df\, f P(f,\theta)$, the {\em force
  density} associated with chains in the $\theta$ direction. \cite{bcol}
This equation can be analyzed in detail for the case where all chains
are directed along the discrete set $\theta_j= 2\pi j/n$ for $0\leq j<n$,
depicted in Fig.~\ref{fig:nstar}.  Consistency with the restriction
to this discrete set can be maintained by assuming that splittings are
always symmetric and outgoing chains make angles of $2\pi/n$ with the
incoming one.  (For $n>8$ one can also have splittings at angles of
$4\pi/n$, and so on, but the simplest case is sufficient for our
purposes.)  To avoid the need for special treatment of the horizontal
directions, we also consider only $n$'s that are not multiples of $4$.

Let $F_j$ be the force density for chains in the direction $\theta = 2\pi
j/n$.  After rescaling of $\lambda$ to unity, the continuum equations
for the $F$'s are
\begin{equation}\label{eqn:nfold}
\cos (j\theta) \, \partial_z F_j + \sin (j\theta) \partial_x F_j
 = -F_j + \frac{1}{2c_1}\left(F_{j-1}+F_{j+1}\right),
\end{equation}
where $\theta\equiv 2\pi/n$, $c_1\equiv\cos\theta$, and indices are taken
modulo $n$.  An analysis of this equation in the slab geometry is
presented in the appendix.  It shows that the lack of solutions to the
linear problem in the 6-fold case is generic: force densities always
diverge when the system is sufficiently deep, even though the
intensities of the chains become exponentially small after many
splittings.

\subsection{Three dimensions}
In 3D, the analogue of the $120^{\circ}$ vertex for splitting and
fusing is the tetrahedral vertex.  Three types of events are possible:
splitting, in which one chain splits into three; fusion, in which
three chains fuse into one; and scattering, in which incoming chains
along two of the tetrahedral directions produce two outgoing chains
along the remaining two directions.  In principle, scattering events
should be more common than fusions, which require terms in the \eqname
equation of order $P^3$.  Scattering events do not change the total
force density in the system, however, and therefore are not sufficient
to regulate the divergence in the linear theory.  Though scattering
events should clearly be included in the \eqname equation for
completeness, fusions are the essential process.

A suitable star of vectors is formed from the 20 vectors pointing to
the faces of an icosahedron.  (The apparently simpler choice of the
star of vectors pointing to the faces of an octahedron turns out to
possess certain symmetries that lead to nongeneric behavior.)  Each
vector on the icosahedral star is a part of two different tetrahedra.
One can assume, for example, that the splitting of a chain occurs
along each tetrahedral set with probability $1/2$.

Choosing the orientation of the icosahedral star to include one vector
in the downward vertical direction, and neglecting scatterings, at
least one line of fixed points can be found.  In this orientation,
there is one vector $\vh$ with $z$-component $s_z=1$, three with
$s_z=\sqrt{5}/3$, six with $s_z=1/3$, and 10 related to these by
inversion.  The fixed points can be parametrized by $p$, with the
densities of chains as follows: $P=p^3/\sqrt{2}$ for all $s_z=1$
chains; $P=p^2/\sqrt{2}$ for all $s_z=\sqrt{5}/3$ chains;
$P=p/\sqrt{2}$ for all $s_z=1/3$ chains; and for all chains related to
these by inversion, $P\rightarrow 1/(2P)$.  This gives for the
``Poisson ratios''
\begin{eqnarray}
\nu_1 & = & \frac{8r_{1+}+2r_{2+}}{2r_{1+}+10r_{2+}+3r_{3+}}; \\
\nu_2 & = & \frac{8r_{1-}+4r_{2-}}{2r_{1-}+20r_{2-}+9r_{3-}}.
\end{eqnarray}
$\nu_1$ varies from $2/3$ at the fixed point, which is outside the
range of stability for a standard, isotropic elasticity theory, to $0$
at large $p$.  $\nu_2$ varies from $16/69$ to $0$ and thus is always
within the stable regime. 

This preliminary treatment of a 3D model suggests that there is
nothing fundamentally different from the 2D models studied above.  The
spectrum and response functions in the vicinity of a fixed point can
be calculated numerically.  Extracting the asymptotic form of the
response, however, is not as straightforward as in the 2D case.
Thorough analyses of the nature of the response and the effects of the
scattering terms are beyond the scope of this work.

\subsection{Long-wavelength theories for a continuum of chain directions}
In Ref.~\cite{bcol}, equations for the stress tensor were developed
for a directed force chain network without the restriction to a
discrete set of directions.  It was shown that for the {\em linear}
\eqname equation, multiplication by $f$ and integration over $f$ could
be carried out to obtain equations for the force densities
$F(\theta)\equiv\int_0^{\infty} df\, f P(f,\theta)$.  These in turn could be
cast in terms of equations for the stress tensor under the assumption
that $F(\theta)$ was nearly isotropic.  This assumption is clearly not
true on length scales of order $\lambda$, but was argued to be true at
large length scales.  In terms of $F$, the following three quantities
were defined:
\begin{eqnarray}
\rho & = & \frac{1}{2}\int_{-\pi}^{\pi}d\Omega\, F(\theta)\,; \nonumber \\
\Jv & = & \frac{1}{2}\int_{-\pi}^{\pi}d\Omega\, F(\theta)\,\nh\,; \\
\tens{\sigma} & = & \int_{-\pi}^{\pi}d\Omega\, F(\theta)\;\nh\otimes\nh, \nonumber
\end{eqnarray}
where $d\Omega=d\theta/2\pi$ as defined above.  Here $\rho$ is the
hydrostatic pressure, $\tens{\sigma}$ is the stress tensor (see
Eq.~(\ref{eqn:stresstensor}), and $\Jv$ is a vector field that
plays a similar mathematical role to the displacement field in
standard elasticity theory.  Under the assumption that $F(\theta)$ is
nearly isotropic, one can express $F(\nh)$ as a linear combintation of
$\rho$, $\nh\cdot\Jv$, and $\nh\cdot\tens{\sigma}\cdot\nh$, then
use Newton's laws and a compatibility condition on $\Jv$ to derive
a partial differential equation for $\tens{\sigma}$.  That equation
turned out to be elliptic, which suggested a response similar to that
of an elastic material.\cite{bcol}

From the results of the present work, it appears that the assumption
of the isotropy of $F(\theta)$ at large length scales was not justified.
Analysis of the linear theory, as discussed in the appendix, shows
that the isotropic fixed point $F=0$ is unstable to perturbations that
favor large $F(\theta)$ for $\theta$ near $\pi/2$, and that the nontrivial
fixed points are generically anisotropic.  Thus the analysis presented
in Ref.~\cite{bcol} should be generalized to the anisotropic case.  It
is not clear whether such a generalization can preserve the close
correspondence between $\Jv$ and the displacement field of
standard elasticity theory.

In addition to the issue of anisotropy, this program encounters
difficulties related to the neglect of the nonlinear terms in the
derivation of the equations for $F(\theta)$.  Naive attempts to carry out
the same program without neglecting the nonlinear terms fail: the
structure of the $\phi_f$ terms in Eq.~(\ref{eqn:master}) make it
important to calculate the full distribution $P(f,\theta)$.  It is not
possible, without making further approximations, to recast the
equations in terms of $F(\theta)$ alone.\cite{ottoprep}

It should also be noted that the implicit passage to the limit of
large length scales corresponds to keeping only terms of order $q$ in
the expressions for $c$ and $D$.  Such a calculation does not pick up
the quadratic dependence of $D$ on $q$.  Thus, in these theories,
whenever the response is hyperbolic the apparent value of $D$ is zero,
correponding to delta-functions that propagate without broadening.
Inclusion of the terms in the expansion of $F$ containing nonzero
wavevectors $\qv$ would be required in order to compute $D$.

\section{Discussion}
\label{sec:conclusion}
\subsection{Numerical simulations}
There is no known algorithm for numerically generating configurations
that are consistent with the \eqname equation.  Simulations previously
reported \cite{bcol} were done on small systems and used a
regularization scheme that is uncontrolled.  The difficulty
encountered in these simulations is that the effects of fusions can
create ``causality'' problems: a chain that gives rise to many others
through multiple splittings may itself be altered by a fusion with one
of its offspring.  For small systems, this problem does not occur and
the pseudo-elasticity theory derived from the linear \eqname equation
appears to give a good description response.\cite{bcol} For large
systems, however, it is an important part of the physics.  The
algorithms used in Ref.~\cite{bcol} for generating DFCN's grind to a
halt when the system size is increased to a few times the length
$\lambda$, due to the generation of infinite loops associated with
fusions.

One approach to avoiding the causality problem would be to neglect
fusions altogether.  To avoid an exponential explosion in the number
of chains (and the total force density), one must then introduce a
lower cutoff on force chain intensities.  All chains with intensity
lower than the cutoff are simply ignored.  Again, this approach may be
reasonable for small systems where the chains being neglected are just
those that were generated in splitting events where one daughter chain
was in nearly the same direction as the parent and the other was very
weak.  For very large systems, however, chains split many times before
reaching the bottom.  At large depths, the number of chains surviving
above the cutoff will eventually decay to zero, so no stress at all
will be transmitted to large distances.  The apparently justifiable
neglect of small forces leads to substantial violations of Newton's
laws.

It appears that the numerical generation DFCN's will require
algorithms for relaxing candidate configurations in a nontrivial way
so as to arrive at networks with statistical properties consistent
with the \eqname equation.

\subsection{Strong vs. weak disorder and the importance of discrete directions}
Hyperbolic systems with weak disorder yield diffusively broadening,
propagating peaks.\cite{bcc} The strong disorder re\-gime has not
previously been accessible.  In one respect, our discrete model
exhibits strong disorder: splitting events cause chains to rapidly
lose memory of the direction of their ancestors.  In another sense,
though, the disorder may be weak: the orientation of the star vectors
exhibits no fluctuations.  In any case, the discrete model may apply
rather directly to the case of a hexagonal array of disks with a
random distribution of vacancies.

We are currently investigating the behavior of the \eqname equation
for more general splitting and fusing functions.\cite{ottoprep} To
make predictions for real granular materials that have not been
carefully prepared to have orientational order, it is crucial to know
whether the hyperbolic response at large depths survives in the
absence of a globally specified set of favored directions.

\subsection{Boundary conditions}
A major unanswered question about the DFCN theory is how to determine
the appropriate boundary conditions associated with a particular
physical situation.  Assume, for example, that we clamp a slab of
granular material between to nominally flat pistons.  How do we decide
whether we are injecting a high density of weak chains or a low
density of strong chains?  The problem is complicated by the fact that
some part of the pressure on the pistons derives from the {\em
  response} to the injected chains, so our boundary condition must be
chosen so as to produce the correct pressure at the associated fixed
point; we cannot simply inject chains corresponding to the desired
loading of the surfaces, though in the slab geometry we can always
scale all force chain intensities by a common factor to produce any
desired total vertical force on the top and bottom boundaries.

\subsection{Anisotropy and gravity}
As one might expect on symmetry grounds, generic fixed point solutions
of the \eqname equation in the slab geometry are anisotropic even
though the equation does not contain intrinsic symmetry breaking
terms, and this anisotropy has a qualitative effect on the response.
The source of the differences in response are traceable to the
inherent nonlinearity of the system.  The response near a given fixed
point cannot be construed as the superposition of the response in an
isotropic system together with a homogeneous anisotropic deformation.

Inclusion of gravity in the DFCN theory may be accomplished by
assuming that every grain is a source of a new vertical chain, which
may or may not instantly fuse with other chains passing through the
same grain.  This would require the addition of a source term to the
\eqname equation that breaks the isotropy of the equation itself, not
just the solutions with anisotropic boundary conditions.  Similar
considerations would apply for systems with an anisotropic fabric
tensor; i.e., systems in which certain directions are favored by an
intrinsic anisotropy in the packing geometry.  Analysis of systems
with intrinsic anisotropy is beyond the scope of this work.

\subsection{Standard elasticity}
Materials described by standard, linear elasticity theory, even if
isotropic, always have Fourier spectra with exponents linear in $q$.
Such theories can produce propagating peaks, as have been observed for
ball-and-spring models with strong anisotropy \cite{gg}. But the peaks
must always broaden linearly, never diffusively.  For classically
stable materials, the peaks will always broaden linearly with depth.
For a semi-infinite slab, there is no parameter in the theory with
dimensions of length, so the response is always a function of $x/z$
only.  Outside the domain of stability, however, the response may be
hyperbolic.  We are currently investigating anisotropic models in this
regime in an effort to determine whether they can describe the
bahavior of DFCN's at very large length scales.\cite{ottoprep}

\subsection{Interpreting experiments}
The emergence of a two-peaked, hyperbolic response at large depths in
the models we have studied is quite remarkable, particularly in the
vertical orientation of the 6-fold star, where the directions of
propagation are not simply related to the star vectors.  In a strongly
scattering system such as this one, where splittings cause large
changes in the directions of force chains, one might naively expect
any peaks to be strongly broadened at large depths, which would lead
to a single centered peak in the response function.  Indeed, this is
what we observe for moderate depths: a system may have a two-peaked
response close to the surface, but those peaks decay or broaden
rapidly, and a single peak emerges.  Surprisingly, in the \dy\uy -model
two peaks that really are associated with hyperbolic response
emerge at even greater depths, and the directions of propagation of
these peaks are generally not the same as those of the original two
peaks near the surface.

Experiments on 2D systems have shown both propagating peaks and single
centered peaks.\cite{behringer,clement,blair} In interpreting these
experiments it is important to note three things: (1) Correlations in
the positions of grains at the surface with the layer just below may
affect the interpretation of the distribution of chain directions
injected at the surface.  It would not be surprising to see structure
in the response close to the surface, possibly even double peaks due
to the immediate splitting of a vertical force chain.  (2) For
well-ordered systems, the splitting length $\lambda$ may be large
enough so that the entire system is in the very shallow regime.  Thus
the two peaks observed are not necessarily indicative of a hyperbolic
response at large scales.  (3) Even for disordered systems, one may
have to probe systems an order of magnitude or more larger than
$\lambda$ before the true large system behavior becomes apparent.
This may be especially difficult because the peaks decrease in
amplitude at large depths, but the fact that diffusively broadening
peaks become cleanly separated at large depths offers some hope that
they could be resolved in future experiments.

\begin{acknowledgement}
  \textbf{Acknowledgements}\\
  We thank J.-P.~Bouchaud and M.~Otto for extensive discussions and
  editorial comments, and R.~Behringer, E.~Cl\'{e}ment, E.~Kolb,
  D.~Levine, G.~Ovarlez, and G.~Reydellet for helpful conversations.
  J.E.S.S. is grateful for the hospitality of the LMDH at
  Jussieu in Paris, where much of this work was carried out.  This
  work was supported by the grant PICS-CNRS~563, NSF Grants
  DMS-98-03305 and PHY-98-70028.
\end{acknowledgement}


\appendix
\section{Failure of the linear theory near the origin}
In this appendix we present an analysis of the linear \eqname equation
for a discrete set of chain directions and symmetric splittings.  The
main purpose is to clarify the nature of the divergence and hence the
failure of the linear theory.

\subsection{The equations and their solutions}
Let $F_j$ be the density of force chains with direction $\theta = 2\pi
j/n$, where $n$ is an integer that is not a multiple of $4$ and
$\theta=0$ corresponds to the positive $z$ direction.  (See
Fig.~\ref{fig:nstar}.) Assume that splittings always occur at an
angle $\theta=2\pi/n$; i.e., when a chain in the $j$ direction splits,
the only option is to produce chains in the $j-1$ and $j+1$
directions.  Eq.~(\ref{eqn:nfold}), reproduced here, governs the force
densities.
\begin{equation}
\cos (j\theta)\, \partial_z F_j + \sin (j\theta)\, \partial_x F_j
 = -F_j + \frac{1}{2c_1}\left(F_{j-1}+F_{j+1}\right).
\end{equation}
We seek solutions to this equation in the slab geometry with a load that
is uniform in the horizontal direction.

Translational invariance ensures that the solution will be uniform in
the horizontal direction, so the $\partial_x$ terms vanish.  Letting
$c_j$ stand for $\cos (j\theta)$, we have
\begin{equation}
\partial_z F_j  = -\frac{1}{c_j}F_j + \frac{1}{2 c_1 c_j}\left(F_{j-1}+F_{j+1}\right).
\end{equation}
In matrix form,
\begin{equation}
\partial_z \Fv = \Mm\cdot\Fv,
\end{equation}
where $\Fv$ is the $n$-dimensional vector with elements $F_j$ and
\begin{equation}\label{eqn:matrixm}
\Mm = \left( \begin{array}{ccccccc}
\frac{1}{c_0} & \frac{1}{2 c_1 c_0} & 0 & \cdots & \ & 0 & \frac{1}{2 c_1 c_0} \\
\frac{1}{2 c_1 c_1} & \frac{1}{c_1} & \frac{1}{2 c_1 c_1} & 0 & \cdots & \ & \ \\
0 & \frac{1}{2 c_1 c_2} & \frac{1}{c_2} & \frac{1}{2 c_1 c_2} & 0 & \cdots & \ \\
\vdots & \ & \ddots & \ddots & \ddots & \ & \ \\
\frac{1}{2 c_1 c_{n-1}} & 0 & \cdots & \ & 0 & \frac{1}{2 c_1 c_{n-1}} & \frac{1}{c_{n-1}}
\end{array}\right).
\end{equation}

$\Mm$ is tridiagonal (except for the entries in the corners), but
{\em not} symmetric.  Its eigenvectors are not orthogonal, nor do they
span the full space.  One can check that both 
\begin{equation}\label{eqn:c2m} 
\Cv^{(2)}\cdot\Mm = 0 
\end{equation} 
and
\begin{equation}\label{eqn:s2m} 
\Xv^{(2)}\cdot\Mm = 0,
\end{equation} 
where $\Cv^{(2)}$ and $\Xv^{(2)}$ are the (row) vectors with
components $c_j^2$ and $c_j \sin (j\theta)$, respectively. 
This implies that $\Cv^{(2)}$ and $\Xv^{(2)}$ are orthogonal to every
eigenvector of $\Mm$ with a nonzero eigenvalue, and therefore that no
matter how those eigenvectors grow or shrink with increasing $z$, the
projection of their sum onto $\Cv^{(2)}$ and $\Xv^{(2)}$ will be
constant determined by the amplitudes of the eigenvectors with
eigenvalue zero. These
projections corresponds to $\sigma_{zz}$ and $\sigma_{xz}$, and their
invariance is a direct consequence of Newton's laws.

The matrix $\Mm$ can be block diagonalized to separate subspaces
that are symmetric or antisymmetric with respect under $j\rightarrow
n-j$.  For present purposes, it is sufficient to consider only the
symmetric subspace.  Let $\Mm_s$ denote the $m\times m$ matrix
corresponding to the symmetric block.  For $n$ odd, we have
$m=(n+1)/2$ and for $n$ even $m=(n/2)+1$.  $\Mm_s$ has two
eigenvalues equal to zero.  One is associated with the eigenvector
$\Cv$ with elements $F_j = c_j$.  The other is associated with a
vector $\Ev_1$ that satisfies $\Mm\cdot \Ev_1 = \Cv$
rather than the usual eigenvalue equation $\Mm\cdot \Ev =
\lambda \Ev$.  One can check that the solution is $\Ev_1 =
b(1,1,1,\cdots) - a \Cv^{(2)}$, with $a = c_1 / (c_1-c_2)$ and $b
= a (1 - c_1)$.

The solution of the differential equations for an arbitrary initial
condition within the symmetric subspace is
\begin{equation}\label{eqn:fgeneral}
\Fv = a_0 \Cv + a_1 \left(\Ev_1 + z \Cv\right) 
          + \sum_{k=2}^{m-1} a_k e^{\lambda_k z}\Ev_k,
\end{equation}
where the $a_k$'s are determined by the boundary conditions.  We
emphasize the following three features of the solutions.  First,
projections onto $\Cv$ are {\em not} conserved as $z$ varies, even
though the eigenvalue associated with $\Cv$ is zero, because $\Cv$
appears also multiplied by $z$.  Second, symmetry under $z\rightarrow
-z$ and $F_j\rightarrow F_{n/2-j}$ guarantees that if $\lambda$ is an
eigenvalue of $\Mm_s$, $-\lambda$ is also, implying that, aside from
the eigenvectors with zero eigenvalues, half of the eigenvectors grow
exponentially with increasing $z$.  Finally, all the $\Ev_k$'s except
$\Ev_1$ must have both positive and negative components, since their
projection onto the positive definite $\Cv^{(2)}$ vanishes.

\subsection{Negative force densities and their origin}
To avoid negative force densities in a deep system, the boundary
values of $\Fv$ must have zero projection on all eigenvectors with
positive eigenvalues.  Since $\sigma_{zz}=\Cv^{(2)}\cdot \vec{F}$,
any boundary condition at $z=0$ that has nonvanishing $\sigma_{zz}$
must have $a_1\neq 0$.  But this implies that the $z \vec{C}$ term in
Eq.~(\ref{eqn:fgeneral}) cannot be avoided.  Since $\vec{C}$ has
negative components, and all the contributions from eigenvectors with
negative eigenvalues decrease with increasing $z$, the solution will
always produce negative force densities at sufficiently large $z$.

The appearance of negative force densities may be counter-intuitive.
After all, the linear equations describe physically allowable
processes that can never generate negative densities of chains and
never yield negative intensities for any chain.  The source of the
negative $F$'s in the solution is a divergence that occurs when the
system is sufficiently deep to allow enough splittings that an
appreciable fraction of the descendants of a downward-pointing chain
are directed upwards.  This causes an exponential explosion in the
number of chains in the system and a consequent divergence of the
force densities.  Even though the intensities of the descendants of a
given chain decrease exponentially with generation number, the sum of
their intensities diverges because in every splitting event from a
chain with intensity $f_1$ to two with $f_2$ and $f_3$ we have
$(f_2+f_3)/f_1=\sec(\theta)>1$.  The effect on the solutions to the
differential equations is roughly analogous to the continuation of
$\sum_{m=0}^{\infty}r^m=1/(1-r)$ beyond its radius of convergence
$r=1$, where it appears that an infinite sum of positive terms gives a
negative result.  In other words, the theory is not well defined for
slabs of thickness larger than a ``persistence length'' for chain
directions.

\subsection{The question of anisotropy at large scales}
The problem of negative densities in the linear theory shows that the
response of the DFCN system can only be properly defined for a
pre-stressed system.  This still leaves open the possibility that the
same linear theory might describe deviations from some uniformly
stressed state.

Here we point out that in the slab geometry with uniform horizontal
loading, the instability of the trivial solution, $F=0$, to the linear
equations leads to anisotropic $F$'s for a broad class of splitting
functions $\psi_{s}$.  For discrete $n$-fold systems in which
symmetric splittings ($\theta_2=\theta_3$) are allowed with uniform
probability for all angles up to $\theta_2=\pi/3$, numerical
diagonalization of the matrix $\Mm$ of Eq.~(\ref{eqn:matrixm})
shows that the most strongly unstable $q=0$ modes correspond to
eigenvectors $E(\theta)$ peaked near $\theta=\pm\pi/2$.  The eigenvalues
grow more unstable and the eigenvectors more sharply peaked as $n$
increases toward the continuum limit.  Analytical arguments have also
been obtained for certain special choices of $\psi_s(\theta_2,\theta_3)$.

The anisotropic nature of this instability casts doubt on the validity
of the conjecture that the DFCN should appear isotropic on large
scales.  The isotropic DFCN does not appear to be an attractor for
any but a perfectly tuned set of initial conditions.

\end{document}